\PassOptionsToPackage{usenames,dvipsnames}{xcolor}
\documentclass[twocolumn,twocolappendix]{aastex631}
\pdfoutput=1 
\usepackage[T1]{fontenc}
\usepackage{ae,aecompl}
\usepackage[utf8]{inputenc}
\usepackage[english]{babel}
\usepackage{amsmath,amstext}
\usepackage{apjfonts}
\usepackage{chngcntr}
\usepackage[figure,figure*]{hypcap}
\usepackage[hang]{footmisc}
\shorttitle{EDEN: Exploring Disks Embedded in N-body simulations}
\shortauthors{Wang et al.}

\usepackage{wasysym}
\usepackage{amsmath,amstext}
\usepackage[T1]{fontenc}
\usepackage{ae,aecompl}
\usepackage[utf8]{inputenc}
\usepackage{apjfonts} 
\usepackage[figure,figure*]{hypcap}
\usepackage{enumitem}
\usepackage[hang]{footmisc}
\allowdisplaybreaks

\begin{document}

\title{EDEN: Exploring Disks Embedded in N-body simulations of Milky-Way-mass halos from Symphony}

\author[0000-0001-8913-626X]{Yunchong Wang}
\affiliation{Kavli Institute for Particle Astrophysics \& Cosmology,  Stanford University, 452 Lomita Mall, Stanford, CA 94305, USA}
\affiliation{SLAC National Accelerator Laboratory, 2575 Sand Hill Road, Menlo Park, CA 94025, USA}
\affiliation{Department of Physics, Stanford University, 382 Via Pueblo Mall, Stanford, CA 94305, USA}

\author[0000-0001-9863-5394]{Philip Mansfield}
\affiliation{Kavli Institute for Particle Astrophysics \& Cosmology,  Stanford University, 452 Lomita Mall, Stanford, CA 94305, USA}
\affiliation{SLAC National Accelerator Laboratory, 2575 Sand Hill Road, Menlo Park, CA 94025, USA}

\author[0000-0002-1182-3825]{Ethan O.~Nadler}
\affiliation{Department of Astronomy \& Astrophysics, University of California, San Diego, La Jolla, CA 92093, USA}
\affiliation{Carnegie Observatories, 813 Santa Barbara Street, Pasadena, CA 91101, USA}
\affiliation{Department of Physics $\&$ Astronomy, University of Southern California, Los Angeles, CA, 90007, USA}

\author[0000-0002-8800-5652]{Elise Darragh-Ford}
\affiliation{Kavli Institute for Particle Astrophysics \& Cosmology,  Stanford University, 452 Lomita Mall, Stanford, CA 94305, USA}
\affiliation{SLAC National Accelerator Laboratory,  2575 Sand Hill Road, Menlo Park, CA 94025, USA}
\affiliation{Department of Physics, Stanford University, 382 Via Pueblo Mall, Stanford, CA 94305, USA}

\author[0000-0003-2229-011X]{Risa H.~Wechsler}
\affiliation{Kavli Institute for Particle Astrophysics \& Cosmology,  Stanford University, 452 Lomita Mall, Stanford, CA 94305, USA}
\affiliation{SLAC National Accelerator Laboratory,  2575 Sand Hill Road, Menlo Park, CA 94025, USA}
\affiliation{Department of Physics, Stanford University, 382 Via Pueblo Mall, Stanford, CA 94305, USA}

\author[0000-0002-5421-3138]{Daneng Yang} \affiliation{Department of Physics and Astronomy, University of California, Riverside, CA 92521, USA}

\author[0000-0002-8421-8597]{Hai-Bo Yu} \affiliation{Department of Physics and Astronomy, University of California, Riverside, CA 92521, USA}

\begin{abstract}
We investigate the impact of galactic disks on the tidal stripping of cold dark matter subhalos within Milky Way (MW)-mass halos ($M_{\rm vir}\sim 10^{12}\mathrm{M_{\astrosun}}$) using a new simulation suite, EDEN. By re-simulating 45 MW-mass zoom-in halos from the N-body Symphony compilation with embedded disk potentials, which evolve according to star formation histories predicted by the  \textsc{UniverseMachine} model, we self-consistently tie disk growth to halo accretion rate and significantly expand the range of disk masses and formation histories studied. We use the particle-tracking-based subhalo finder \textsc{Symfind} to enhance the robustness of subhalo tracking. We find that disks near the median disk-to-halo mass ratio of our sample ($M_{\ast, \rm Disk}/M_{\rm vir, host} = 2\%$) reduce subhalo peak mass functions within 100 kpc by about $10\%$ for peak masses above $ 10^8\mathrm{M_{\astrosun}}$. Heavier, MW/M31-like disks ($M_{\ast, \rm Disk}/M_{\rm vir, host} \gtrsim 5\%$) lead to a reduction of more than $40\%$. Subhalo abundance suppression is more pronounced near halo centers, with particularly enhanced stripping for subhalos accreted over 8 Gyr ago on orbits with pericenters < 100 kpc. Suppression is further amplified when disk mass is increased within fixed halo and disk assembly histories. In all cases, the suppression we measure should be interpreted as stripping below the mass resolution limit rather than complete subhalo disruption. This study reshapes our understanding of the MW's impact on its satellites, suggesting it strips subhalos more efficiently than typical MW-mass galaxies due to its larger disk-to-halo mass ratio and earlier disk formation.
\end{abstract}
\keywords{ \href{http://astrothesaurus.org/uat/595}{Galaxy formation (595)},
 \href{http://astrothesaurus.org/uat/1083}{N-body simulations (1083)},
 \href{http://astrothesaurus.org/uat/1880}{Galaxy dark matter halos (1880)}}

\section{Introduction} 
\label{sec:intro}

The abundance, spatial distribution, rotation velocity, and sizes of low-mass galaxies (stellar mass $M_{\ast}\lesssim 10^9\mathrm{M_{\astrosun}}$) in the Milky-Way (MW) and similar galaxies in the local Universe ($D\lesssim 100$ Mpc) can inform the fundamental physics of a range of models for galaxy formation and the nature of dark matter (see \citealt{2017ARA&A..55..343B,2022NatAs...6..897S} for reviews). A number of physical processes could potentially impact the properties of low-mass galaxies, including cosmic reionization photoionizing gas in low-mass (sub)halos and shutting down their star formation completely since $z\gtrsim 6$~\citep{2000ApJ...539..517B,2002ApJ...572L..23S}; stellar feedback preventing further star formation and creating dark matter cores that make subhalos more prone to tidal stripping~\citep{2012MNRAS.422.1231G,2012MNRAS.421.3464P,2012ApJ...761...71Z,2021MNRAS.502..621J}; alternative dark matter models suppressing the non-linear power spectrum curtailing structure growth on small scales~(e.g., \citealt{2001ApJ...556...93B,2012MNRAS.424..684S,2013JCAP...03..014A,2014MNRAS.439..300L,2016MNRAS.460.1399V,2021ApJ...920L..11N}); the specific assembly history of the host galaxy such as a recently infalling Large Magellanic Cloud~\citep[LMC, ][]{2020ApJ...893...48N} can all lead to significant changes in the abundance and radial distributions of satellites.  

Here, we systematically investigate a major contributing effect that could significantly alter the predicted abundance and radial distributions of satellites and their subhalos: the enhanced mass loss due to the strong tidal forces from the baryonic mass content (stars and gas) of their host galaxies. This effect was originally brought forward as a potential resolution to the small-scale challenges of the $\Lambda$CDM cosmological model, i.e. the missing satellite problem~\citep{1999ApJ...522...82K,1999MNRAS.310.1147M} and the too-big-to-fail problem~(TBTF, \citealt{2011MNRAS.415L..40B,2012MNRAS.422.1203B,2014MNRAS.440.3511T}), that could mitigate the order-of-magnitude gap between predicted abundances of subhalos in dark-matter-only (DMO) simulations~\citep[e.g.,][]{2008MNRAS.391.1685S,2009Sci...325..970K,2009MNRAS.398L..21S,2014MNRAS.438.2578G,2015ApJ...810...21M,2016ApJ...818...10G,2023ApJ...945..159N} and the observed number of low-mass satellites in the Milky Way \citep[e.g.,][]{2012AJ....144....4M}.

The enhanced mass loss effect from the host galaxy is primarily studied using pairs of hydrodynamic simulations and DMO counterparts starting out from the same initial conditions, focusing on satellites around MW-mass halos~\citep{2013MNRAS.431.1366S,2014ApJ...786...87B,2015MNRAS.448.2941S,2016ApJ...827L..23W,2016MNRAS.458.1559Z,2017MNRAS.469.1997D,2017MNRAS.467.4383S,2017MNRAS.471.1709G,2018MNRAS.480.1322G,2020MNRAS.491.1471S,2023MNRAS.523..428B,2024ApJ...964..123J}. It is revealed that significant suppression of satellite and subhalo abundances occurs, especially near host centers for subhalos with smaller pericenters and more radially-biased velocities due to the addition of baryons (stars and gas) compared to the DMO case. Since this effect is primarily due to the central galaxy's mass concentration, it is, in principle, agnostic towards the specific treatment of star formation and feedback physics in the hydrodynamic simulations. 

To further isolate the pure gravitational effect of the central galaxy, several simulation works further combined analytic disk potentials with DMO zoom-in simulations~\citep{2010ApJ...709.1138D,2017MNRAS.465L..59E,2017MNRAS.471.1709G,2019MNRAS.487.4409K}. These explorations also discovered similar levels of subhalo abundance suppression as hydrodynamic simulations when compared to their DMO counterparts. These studies interpreted reduced subhalo abundances as complete physical disruption and thus concluded that the central galaxy alone could potentially solve the missing satellites and TBTF problems.\footnote{The latter problem is addressed because heavier subhalos experience stronger dynamical friction and sink more quickly toward halo centers, where the disk effects are strongest.} However, we will argue that reduced subhalo abundances above a peak (or present-day) mass threshold is \emph{not} evidence of physical disruption; rather, it indicates enhanced mass loss which is in agreement with recent semi-analytic studies~\citep{2022MNRAS.509.2624G}. Furthermore, previous studies often implicitly assume that satellite galaxies hosted by subhalos would ``disrupt'' along with the subhalo itself, which is also an oversimplification as ``orphan'' galaxies out-surviving their subhalos are required to match observations~\citep{2019MNRAS.488.3143B} and depend on the specific resolution of the simulation used~\citep{2024ApJ...970..178M}.  Thus, we are motivated to revisit how disk potentials affect subhalo populations in both the MW and ``typical'' MW-mass halos. 

The MW lives in a halo that has a virial mass of $M_{\rm vir}\sim 10^{12}\mathrm{M_{\astrosun}}$~\citep{2016ARA&A..54..529B,2017MNRAS.465...76M,2018A&A...616A..12G,2019A&A...621A..56P,2019ApJ...873..118W,2020MNRAS.494.4291C,Labini_2023}. All the DMO-embedded disk simulations mentioned above focused on MW-mass halos with $M_{\rm vir}\sim 10^{12}\mathrm{M_{\astrosun}}$ and embedded massive disk potentials similar to our Galaxy~\citep[$M_{\rm d}\gtrsim 5\times 10^{10}\mathrm{M_{\astrosun}}$, ][]{2015ApJ...806...96L,2016ARA&A..54..529B,2017MNRAS.465...76M,2020MNRAS.494.4291C}. Furthermore, in simulations such as \citet{2010ApJ...709.1138D} (fixed disk-to-halo mass ratio) and Phat ELVIS (abundance matching $M_{\ast}$ to $M_{\rm vir}$, \citealp{2013ApJ...770...57B}), the simplistic disk growth models adopted do not account for diversity in halo assembly and stellar mass growth in a correlated manner. Although \citet{2017MNRAS.471.1709G} used more physical disk formation histories from their hydrodynamic runs, their sample size was limited to only two MW-mass hosts. Therefore, a large parameter space of disk-to-halo mass ratios and disk growth histories is yet to be explored for MW-mass halos in terms of their subhalo abundance suppression effects.

In this work, we aim to fill this gap by better quantifying subhalo abundances over a wider range of disk masses and disk formation histories in MW-mass halos. We present the EDEN (Exploring Disks Embedded in N-body) simulations where we re-simulate 45 $M_{\rm vir}\sim 10^{12}\mathrm{M_{\astrosun}}$ MW-mass halos from the Symphony~\citep{2023ApJ...945..159N} zoom-in simulation compilation with embedded analytic disk potentials. Two technical highlights of this work are that: i) EDEN evolves disk growth according to the empirical galaxy--halo connection model \textsc{UniverseMachine}~\citep[UM hereafter;][]{2019MNRAS.488.3143B,2021ApJ...915..116W} that explicitly correlates the star formation history with the halo assembly history; ii) EDEN uses the state-of-the-art particle-tracking subhalo finder \textsc{Symfind}~\citep{2024ApJ...970..178M} to enable more robust subhalo tracking.  We investigate the reduction in the abundance of subhalos as a function of host mass, subhalo mass, 3D and pericenter distances to their hosts, and infall times. As the disk mass has a dominant impact on subhalo mass loss rates over other factors like disk radius, scale height, and density profiles~\citep{2017MNRAS.471.1709G,2022MNRAS.509.2624G}, we further re-simulate nine of our 45 zoom-in halos with $\times 2.5$ heavier disk potentials while keeping the form of their disk growth histories fixed. This subset of higher-disk-mass simulations, along with the main set of EDEN simulations, provides good sampling of subhalo abundance suppression over the full range of central galaxy stellar masses for typical $M_{\rm vir}\sim 10^{12}\mathrm{M_{\astrosun}}$ MW-mass halos. 

By more realistically modeling subhalos' tidal evolution compared to DMO simulations, EDEN also provides the groundwork for a better understanding of how halo assembly drives satellite radial distributions and radial quenching trends. In the 101 observed MW-mass hosts from the Satellite Around Galactic Analogs (SAGA) Survey~\citep{2017ApJ...847....4G,2021ApJ...907...85M,2024ApJ...976..117M,2024ApJ...976..118G,2024ApJ...976..119W}, the average satellite radial distribution is less concentrated than the typical NFW~\citep{1996ApJ...462..563N} host dark matter density profile~\citep{2024ApJ...976..117M}. The UM-SAGA model~\citep{2024ApJ...976..119W} fit to SAGA satellite data also finds it hard to produce quenched satellite radial distributions as concentrated as the observed satellites. Both of these findings point to the potential effect of the baryonic disk significantly affecting satellite abundance radial distributions. However, in a smaller set of Local Volume~\citep[ELVES, ][]{2021ApJ...922..267C,2022ApJ...927...44C,2023ApJ...949...94G,2023ApJ...956....6D} hosts that cover a broader host mass range than SAGA, \citet{2020ApJ...902..124C} found that their satellites have a more concentrated radial distribution than many DMO and hydrodynamic simulations, although being largely consistent with the top $25\%$ concentration hosts in SAGA~\citep{2024ApJ...976..117M}. This diversity in MW-mass host satellite radial distributions bolsters our emphasis on modeling subhalo populations in a way that accounts for the diversity in disk masses and formation histories, truly placing satellites in our Galaxy~\citep{2012AJ....144....4M,2014ApJ...796...91B,2014ApJ...789..147W,2015ApJ...804..136W,2023ApJ...956...86S} in the cosmological context of more diverse MW-mass host environments surveyed in ELVES and SAGA.

This paper is structured as follows: in Section~\ref{sec:method}, we introduce the MW-mass halos from Symphony and how we embed disk potentials into these zoom-in simulations; in Section~\ref{sec:discussion:disruption}, we review the (mis)concept of subhalo ``disruption'' and discuss why we refrain from interpreting our results with such a term in this work; in Section~\ref{sec:results} we present key results including the effect of disk-to-halo mass ratio on subhalo abundance, subhalo peak mass functions, 3D and pericentric radial distributions, and subhalo infall time distributions; in Section~\ref{sec:discussion} we discuss the various caveats of EDEN, as well as how to interpret our results in broader theoretical and observational contexts; in Section~\ref{sec:conclusion} we summarize our findings.  We adopt a flat $\Lambda$CDM cosmology with $H_0=70\,\mathrm{km\,s^{-1}\,Mpc^{-1}}$, $\Omega_{\rm m} = 0.286$, $\Omega_\Lambda = 0.714$, $n_s=0.96$, and $\sigma_8 = 0.82$.

\section{Methodology} 
\label{sec:method}

Here, we present the methodology for embedding disk potentials into the SymphonyMilkyWay DMO simulations. Section~\ref{sec:method:sim} summarizes basic properties of the Symphony compilation~\citep{2023ApJ...945..159N}; Section~\ref{sec:method:UM} introduces how we apply the galaxy--halo model UM to Symphony and retrieve the stellar mass growth histories of the MW central galaxies. Section~\ref{sec:method:disk} summarizes the disk density components and how we embed them into EDEN host halos at $z=3$; in Section~\ref{sec:method:highmass} we introduce the nine-host subsample that is re-simulated with $2.5\times$ higher disk mass.

\subsection{SymphonyMilkyWay dark-matter-only zoom-ins}
\label{sec:method:sim}

\begin{figure*}
    \centering
	\includegraphics[width=2\columnwidth]{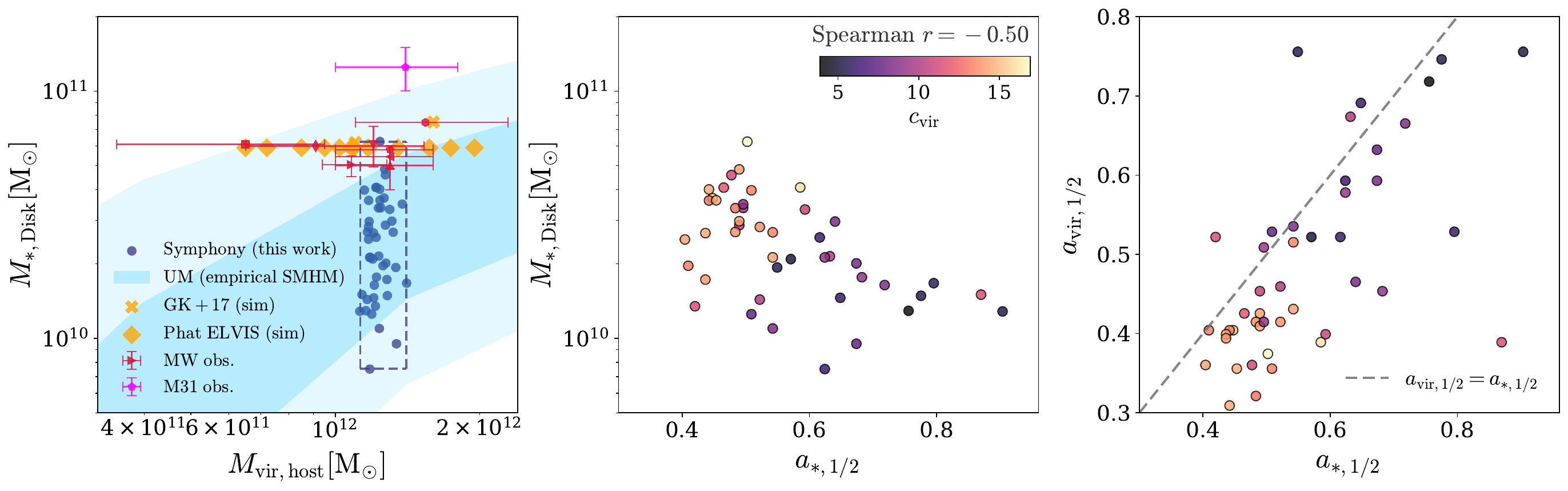}
    \caption{Properties of EDEN host systems. {\it Left}: Stellar mass--halo mass (SMHM) relation of the SymphonyMilkyWay hosts (blue dots, DMO) whose stellar masses are obtained using \textsc{UniverseMachine} (UM). The blue-shaded regions are the $68\%$ and $95\%$ distributions of the global UM SMHM relation derived in the parent cosmological simulation of Symphony, Chinchilla c125-2048. The MW observations (red error bars) consist of stellar masses from \citet{2015ApJ...806...96L} (down triangle), \citet{2016ARA&A..54..529B} (up triangle), \citet{2017MNRAS.465...76M} (left triangle), \citet{2020MNRAS.494.4291C} (right triangle) and halo masses from ~\citet{2016ARA&A..54..529B,2017MNRAS.465...76M,2020MNRAS.494.4291C,2018A&A...616A..12G} (diamond), \citet{2019A&A...621A..56P} (pentagon), \citet{2019ApJ...873..118W} (hexagon), \citet{Labini_2023} (square). The magenta point shows M31 halo mass~\citep{2010MNRAS.406..264W} and stellar mass~\citep{2012A&A...546A...4T} measurements. The observations indicate that the MW and M31 are $\gtrsim 1\sigma$ above the typical stellar mass for an average $M_{\rm vir} \sim 10^{12}\mathrm{M_{\astrosun}}$ halo. Wherever only the halo or stellar mass of the MW is provided, we assign an arbitrary value to the other missing mass for clarity. The orange markers show previous embedded disk DMO simulations from \citet{2017MNRAS.471.1709G} (crosses) and Phat ELVIS~\citep[diamonds,][]{2019MNRAS.487.4409K}, which all target MW-like heavy disk systems. 
    {\it Middle}: UM-predicted stellar mass and stellar half-mass formation scale factor for the SymphonyMilkyWay hosts. Heavier disks also form earlier on average, with a Spearman $r_S=-0.50$. The color bar shows the halo concentration of the MW-mass hosts, with earlier star-forming disks living in more concentrated halos.
    {\it Right}: Stellar versus halo half-mass scale factor for the 45 SymphonyMilkyWay hosts. Most of the MW halos have stellar masses that form later than their halo masses (the dashed line indicates a one-to-one relation). The color map is the same as the middle panel for halo concentration.}
    \label{fig:smhm}
\end{figure*}

The Symphony compilation~\citep{2023ApJ...945..159N} is a set of cosmological zoom-in DMO simulations that focuses on 262 host halos across four decades in halo mass, covering Large-Magellanic-Cloud-mass (LMC for short, $M_{\rm vir}\sim 10^{11}\mathrm{M_{\astrosun}}$), MW-mass ($M_{\rm vir}\sim 10^{12}\mathrm{M_{\astrosun}}$), Group-mass ($M_{\rm vir}\sim 10^{13}\mathrm{M_{\astrosun}}$) and Cluster-mass ($M_{\rm vir}\gtrsim 10^{14.5}\mathrm{M_{\astrosun}}$) halos. The zoom-in host halos are mostly isolated, and the simulations have, in principle, enough resolution to resolve all dark matter halos that host observable satellite galaxies in the Universe (down to $M_{\rm vir, sub}\sim 1.5\times 10^7\mathrm{M_{\astrosun}}$).

In this work, we focus on the 45 MW-mass host halos from Symphony. Originally, these halos were selected from the Chinchilla c125-2048 cosmological simulation~\citep{2015ApJ...810...21M} with masses $M_{\rm vir}/\mathrm{M_{\astrosun}}\in 10^{12.1\pm 0.03}$. Resolution is increased within `zoomed-in' Lagrangian regions that contain particles within $\sim 10 R_{\rm vir}$ of the MW hosts at $z=0$. The initial conditions for the zoom-in regions are generated with the public code \textsc{Music}~\citep{2011MNRAS.415.2101H}.\footnote{\url{https://www-n.oca.eu/ohahn/MUSIC/}} Halos were then re-simulated from $z=99$ using \textsc{Gadget-2}~\citep{2005MNRAS.364.1105S}\footnote{\url{https://wwwmpa.mpa-garching.mpg.de/gadget/}}. All the simulations in this suite have particle masses of $m_{\rm DM}=4\times 10^5 \mathrm{M_{\astrosun}}$ and adopt a comoving Plummer-equivalent force softening scale of $\epsilon = 243$ pc in the high-resolution zoom-in region. This mass resolution is equivalent to having $8192^3$ particles in the 125  $h^{-1}$ Mpc parent box c125-2048. In EDEN, we re-simulate these systems with the same particle resolution and force softening while embedding disk potentials (see Section \ref{sec:method:disk}).

Halos in Symphony were identified using the phase-space halo finder \textsc{Rockstar} \citep{2013ApJ...762..109B}.\footnote{\url{https://bitbucket.org/gfcstanford/rockstar/src/main/}} We use the \citet{1998ApJ...495...80B} virial overdensity definition, which is equivalent to an overdensity of $\Delta_c = 99.2$ at $z=0$. Halo merger trees are constructed using \textsc{consistent-trees}~\citep{2013ApJ...763...18B}.\footnote{\url{https://bitbucket.org/pbehroozi/consistent-trees/src/main/}} We use these \textsc{consistent-trees} catalogs to track subhalos before they cross into their hosts' virial radii ($R_{\rm vir, host}$). We use the \textsc{Symfind}~\citep{2024ApJ...970..178M} particle-tracking subhalo finder after subhalo infall for more reliable subhalo tracking, which is a major improvement of this work over previous non-particle-tracking halo finders used to analyze embedded-disk simulations (e.g., AHF, ~\citealt{2009ApJS..182..608K}, for \citealt{2017MNRAS.471.1709G} and \textsc{Rockstar} for \citealt{2019MNRAS.487.4409K,2023ApJ...945..159N}). All halo catalogs and merger trees for Symphony, including \textsc{Symfind} subhalo catalogs, are publicly available at \url{https://web.stanford.edu/group/gfc/symphony/}. The EDEN halo catalogs (\url{https://web.stanford.edu/group/gfc/gfcsims/build/html/eden_overview.html}) can be accessed as part of the Symphony data products and its simulation code repository (\url{https://github.com/Bulk826R/EDEN}) is publicly available for download. 

\subsection{UniverseMachine star formation histories}
\label{sec:method:UM}

We use the UM-predicted stellar mass histories of MW host galaxies from Symphony~\citep{2023ApJ...945..159N}. UM is a flexible empirical galaxy--halo connection model that predicts galaxy star formation rates (SFRs) given their host halo's mass and assembly history. It is statistically constrained by observational data over a wide range of masses and redshifts ($0<z<8$), including stellar mass functions, galaxy quenched fractions, cosmic SFR densities, UV luminosity functions, and galaxy clustering. The latest version of UM, UM-SAGA, is additionally constrained by low-mass galaxy observations ($M_{\ast}\lesssim 10^9\mathrm{M_{\astrosun}}$) from the SAGA Survey and the Sloan Digital Sky Survey~\citep[SDSS,][]{2012ApJ...757...85G}.  Here, MW stellar mass histories are derived using the zoom-in UM application method described in \citet{2021ApJ...915..116W}, which uses the UM DR1 model~\citep{2019MNRAS.488.3143B} and joins together all 45 DMO zoom-in simulations for unbiased halo accretion rate ranking. We do not expect the choice to use the UM DR1 model vs. UM-SAGA to have any impact, since here we 
focus only on modeling the MW-mass host galaxies ($M_{\ast}\gtrsim 10^{10}\mathrm{M_{\astrosun}}$), and the UM-SAGA update does not impact MW-mass galaxies (Fig. 5 in \citealt{2024ApJ...976..119W}).

\begin{figure*}
    \centering
	\includegraphics[width=2\columnwidth]{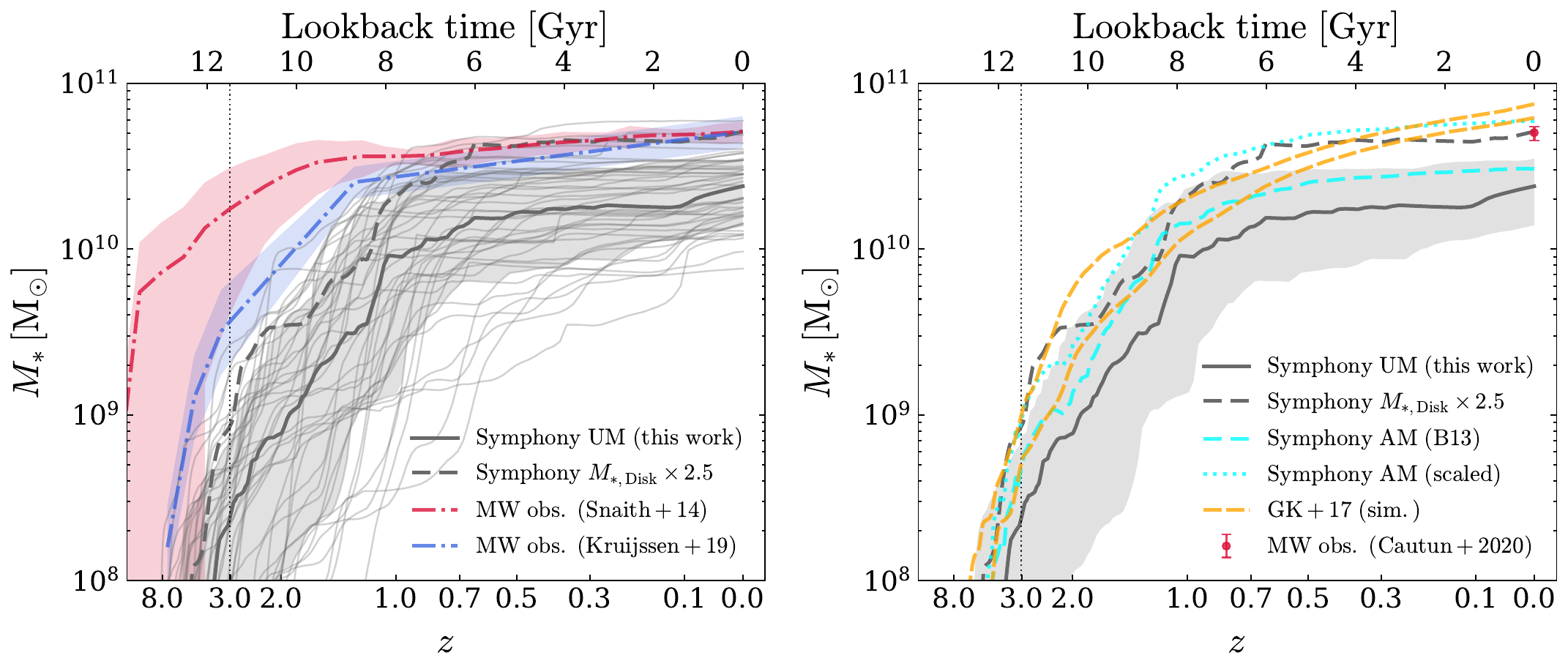}
    \caption{Stellar mass history of the disk in MW-mass systems. The median history of the 45 SymphonyMilkyWay hosts (DMO) with UM-predicted stellar mass histories is shown (thick solid grey; shaded band indicates the [$16\%$, $84\%$] distribution). The median history of nine halos with $\times 2.5$ higher disk masses, which are chosen randomly from the EDEN fiducial sample, is shown as dashed grey (Symphony $M_{\ast, \rm Disk} \times 2.5$); this sample is more consistent than the full Symphony suite with MW observations (left) and previous embedded disk simulations (right). The inferred MW SFHs in the left panel are from \citet{2017MNRAS.471.1709G,2019MNRAS.487.4409K}, and the $z=0$ $M_{\ast}$ of the MW is from \citet{2020MNRAS.494.4291C}. Vertical dotted lines mark $z=3$, when the disks are initialized. {\it Left panel:} The thin gray lines indicate each individual Symphony host. {\it Right panel}: Our simulations are compared stellar mass histories in previous work, including two MW hosts in \citet{2017MNRAS.471.1709G} (dashed orange curves), and Symphony hosts using abundance matching ~\citep[AM, ][]{2013ApJ...770...57B} (dashed turquoise). Phat-ELVIS~\citep{2019MNRAS.487.4409K} used the same AM model; the dotted turquoise curve shows Symphony AM scaled to their final mass of $M_{\ast}=5.9\times 10^{10}\mathrm{M_{\astrosun}}$.}
    \label{fig:sfh}
\end{figure*}

In the left panel of Fig.\ref{fig:smhm}, we show the MW host halo masses ($M_{\rm vir, host}$), the UM-predicted stellar (disk) masses for the MW galaxy ($M_{\ast, \rm Disk}$), the stellar half-mass formation scale factor ($a_{\ast, 1/2}$), and the halo half-mass formation scale factor ($a_{\rm vir, 1/2}$). We assume that the three-component disk potential (Section~\ref{sec:method:disk}) contains the total stellar mass predicted by UM, and we use $M_{\ast}$ and $M_{\ast, \rm Disk}$ interchangeably. Compared to the latest observational constraints on the MW's halo mass~\citep{2016ARA&A..54..529B,2017MNRAS.465...76M,2018A&A...616A..12G,2019A&A...621A..56P,2019ApJ...873..118W,2020MNRAS.494.4291C,Labini_2023}, the SymphonyMilkyWay hosts have quite representative virial masses (mean $M_{\rm vir}\sim 10^{12.1}\mathrm{M_{\astrosun}}$). However, the MW has an atypically large ($\gtrsim 1\sigma$) stellar mass~\citep{2015ApJ...806...96L,2016ARA&A..54..529B,2017MNRAS.465...76M,2020MNRAS.494.4291C} for a halo of $M_{\rm vir} \sim 10^{12}\mathrm{M_{\astrosun}}$, and is larger than the average empirical SMHM relation from UM~\citep{2021ApJ...915..116W}. We also show observational measurements of the halo mass~\citep{2010MNRAS.406..264W} and stellar mass~\citep{2012A&A...546A...4T} of M31, which has even higher stellar-to-halo mass ratio. Previous embedded disk DMO simulations marked in orange from \citet{2017MNRAS.471.1709G} and \citet{2019MNRAS.487.4409K} all focused on MW-like high-disk-mass systems ($\gtrsim1\sigma$ up scatter) for MW-mass halos around $M_{\rm vir}\sim 10^{12}\mathrm{M_{\astrosun}}$. Therefore, a large part of the $M_{\ast, \rm Disk}/M_{\rm vir, host}$ parameter space that the majority of MW-mass galaxies find themselves in has not yet been explored. EDEN aims to fill this important gap and provide a more comprehensive understanding of disk effects on subhalo abundances in MW-mass hosts with a range of galaxy formation histories.

Besides covering a much broader range of disk-to-halo mass ratios than previous literature, EDEN also accounts for the intrinsic scatter in disk formation histories due to the diversity in halo assembly. This is naturally achieved with the UM-predicted stellar mass growth histories of the disks, which include a correlation between galaxy and halo growth. In Fig.~\ref{fig:smhm}, we also show the wide distribution of star formation half mass scale ($a_{\ast, 1/2}$) ranging from 0.4 to 0.9, which anti-correlates with stellar mass and halo concentration $c_{\rm vir}$ (middle panel), and are mostly later than their halo half-mass formation time scale (right panel). Disks with larger stellar masses also form earlier, giving them more time to tidally strip subhalos historically. 

In Fig.~\ref{fig:sfh}, we show the UM-predicted stellar mass growth histories for each EDEN MW-mass host. We compare the host disk formation histories to observational constraints on the MW~\citep{2014ApJ...781L..31S,2019MNRAS.486.3180K} in the left panel. The MW stellar mass growth histories are inferred from the chemical abundances of disk stars~\citep[red, ][]{2014ApJ...781L..31S} and the globular cluster age-metallicity relation~\citep[blue, ][]{2019MNRAS.486.3180K}.  There are differences between the median EDEN star formation history and that of our Galaxy. This is mainly caused by the excessive early (lookback time $\gtrsim 9$ Gyrs) star formation in the real MW, making it more massive and earlier-forming than most EDEN hosts. In the most recent nine Gyrs, EDEN hosts formed $M_{\ast}=10^{10.33}\mathrm{M_{\astrosun}}$ median stellar mass, which is consistent with the MW ($M_{\ast}=10^{10.21}\mathrm{M_{\astrosun}}$ in \citealt{2014ApJ...781L..31S} and $M_{\ast}=10^{10.48}\mathrm{M_{\astrosun}}$ in \citealt{2019MNRAS.486.3180K}). A few of the EDEN hosts have early star formation similar to the MW due to intrinsic scatter in halo assembly, in line with recent findings in hydrodynamic simulations~\citep{2024ApJ...962...84S}. Indeed, the most massive disks predicted by UM are also in halos with the highest concentration ($c_{\rm vir}$) and early halo assembly ($a_{\rm vir, 1/2}$) However, the majority of EDEN hosts have lower disk masses and disks that form later than the MW. This implies they are appropriate for comparison with representative samples of MW-mass halos but are not ideal for comparing directly to MW satellite observations. 

In the right panel, we compare EDEN histories to those adopted by previous embedded disk DMO simulations. We retrieve stellar mass histories of the two halos simulated in \citet{2017MNRAS.471.1709G} from the FIRE-2 public data release~\citep{2023ApJS..265...44W}. We also apply the abundance matching (AM) model~\citep{2013ApJ...770...57B} used by Phat ELVIS~\citep{2019MNRAS.487.4409K} to the SymphonyMilkyWay hosts to demonstrate the effects of modeling galaxy SFR only based on halo mass (removing any correlations with halo assembly). The differences between the AM and UM $M_{\ast}$ come from the slightly lower median $M_{\ast}$ at $M_{\rm vir, host}\sim 10^{12}\mathrm{M_{\astrosun}}$ in UM than the AM and that the individual AM $M_{\ast}$ values are placed at the population mean given their $M_{\rm vir, host}$ without resampling the intrinsic scatter. We also show a `scaled' version of the Symphony AM model stellar mass growth histories that have their median $M_{\ast}$ normalized to the $M_{\ast,\rm Disk}=5.9\time 10^{10}\mathrm{M_{\astrosun}}$ value adopted in Phat ELVIS. Compared to either FIRE-2 or Phat ELVIS, UM stellar mass growth histories cover a much broader range than previously adopted methods and especially better sample the low-$M_{\ast}$ and late-forming MW central galaxies (grey band). This crucial aspect of the present work helps us better place the MW's disk in the context of a broader range of MW-mass halos.

These comparisons between disk growth models highlight the impact of employing UM stellar mass histories for EDEN: not only is the real MW biased towards higher disk masses and earlier star formation, but previous theoretical work has also left the wide range of disk masses and formation histories unexplored due to specific modeling choices. Therefore, EDEN, with its wider coverage of disk masses and formation histories,  should provide a more comprehensive view of subhalo abundance in MW-mass halos.

\subsection{Embedded disk potential setup}
\label{sec:method:disk}

We initialize analytic disk potentials in the $z=3$ snapshot of the SymphonyMilkyWay DMO simulations~\citep[similar to ][]{2017MNRAS.471.1709G,2019MNRAS.487.4409K}. We restart the simulations from $z=3$ and evolve the total equivalent mass of the disk potential in each host following their UM-predicted stellar mass histories (Fig.~\ref{fig:sfh}). We concatenate the output snapshots from the embedded disk simulations at $z\leqslant 3$ with the $z>3$ snapshots in their DMO counterparts for halo finding and merger tree construction. 

The disk potential has three structural components:
\begin{equation}
\label{eq:potential}
    \Phi(R,Z) = \Phi_{\ast}(R,Z)+\Phi_{g}(R,Z)+\Phi_{b}(r),
\end{equation}
where $\Phi_{\ast}$ is the axis-symmetric stellar disk, $\Phi_{\rm g}$ is the axis-symmetric gaseous disk, and $\Phi_{\rm b}$ is the spherical bulge. Numerical tests in \citet{2017MNRAS.471.1709G,2022MNRAS.509.2624G,2024MNRAS.527.8841S} have shown that the disk mass is by far the most decisive factor  in affecting subhalo abundances, hence our specific choices of disk geometry are sub-dominant compared to the disk mass (see discussions in Section~\ref{sec:discussion:disk}). We choose the symmetry axis of the stellar and gaseous disks to be both {\em fixed} and aligned with the MW host halo's spin at $z=0$ throughout the simulation. In the following, we refer to the combined potential $\Phi(R,Z)$ when stating `disk potential'.  

For the density profile of the stellar ($\Phi_{\ast}$) and gaseous ($\Phi_{g}$) disks, we assume that they both follow the commonly used Miyamoto--Nagai potential~\citep{1975PASJ...27..533M}:
\begin{equation}
    \Phi_d(R, Z)=-\frac{G M_d}{\sqrt{\left(\sqrt{h_d^2+Z^2}+R_d\right)^2+R^2}}.
\end{equation}
Here, $G$ is the gravitational constant, $M_{d}$ is the disk mass for `stellar' ($0.60M_{\ast}$) or `gaseous' ($0.27M_{\ast}$) disk, $R_d$ is the disk scale radius (2.5 kpc for stellar and 7 kpc for gaseous at $z=0$), and $h_d$ is the disk scale height (0.35 kpc for stellar disk and 0.084 kpc for gaseous at $z=0$). The gas fraction is consistent with the latest empirical constraints on the HI to stellar mass ratio~\citep{2020MNRAS.496.1124P}, which yields a gas-to-mass ratio of $\sim 30\%$ at $z>2$ and $\sim15\%$ at $z<1$ for MW-mass halos (Fig. 1 in \citealt{2020MNRAS.496.1124P}). As found in \citet{2017MNRAS.471.1709G}, the inclusion of a gas disk potential that follows the gas disk growth in the hydrodynamic run is less important for determining subhalo abundances than the mass of the stellar disk. Motivated by this finding, we assume a fixed fraction of the UM-predicted stellar mass to be in the `gaseous' disk from $z=3$ to $z=0$. We re-simulate three EDEN hosts in Appendix~\ref{sec:app:gas} with gas disks growing more realistically following the predictions of \textsc{NeutralUniverseMachine}~\citep{2023ApJ...955...57G}, an empirical gas--halo connection model calibrated on UM~\citep{2019MNRAS.488.3143B}. We find that the gas disk potential is indeed less consequential than the mass normalization of the stellar disk, in line with \citet{2017MNRAS.471.1709G}. 

As for the disk size, the stellar disk scale radius is consistent with the MW's thin ($2.0\pm 0.2$ kpc) and thick ($2.6\pm0.5$ kpc) stellar disk constraints~\citep{2016ARA&A..54..529B}. The MW gaseous disk is less constrained with typically $\sim 2\times$ the radius of the stellar disk~\citep{2016ARA&A..54..529B,2019A&A...621A..56P}, indicating our assumption is reasonable. Our choice of the stellar disk scale height is representative of MW measurements that span from 140 to 430 pc using different stellar populations~\citep{2015ApJ...814...13M}, while the gaseous disk scale height assumed is consistent with cold CO disk measurements of $\lesssim 100$ pc~\citep{2015ARA&A..53..583H}. In addition to the CO disk, there is also a thin and a thick cold HI disk, both having radially varying scale heights ($h>200$ pc, \citealt{2016A&A...587L...6V}), and we do not model this subtlety in this work. 

Since the spherical bulge ($\Phi_{\rm b}$) only contains $\sim10\%$ of the total disk mass, it is unlikely that our choice of its density profile would significantly change our results. Hence, without loss of generality, we assume a Hernquist potential for the bulge profile~\citep{1990ApJ...356..359H}:
\begin{equation}
    \Phi_{b} (r) = -\frac{GM_{b}}{r+r_{b}}.
\end{equation}
Here, $M_{b}$ is the bulge mass ($0.13M_{\ast}$) and $r_{b}$ is the bulge scale radius (0.5 kpc). The stellar bulge mass measurement of the MW  has large observational uncertainties~\citep{2012A&A...538A.106R,2013JCAP...07..016N,2013PASJ...65..118S,2015MNRAS.448..713P,2016A&A...587L...6V,2017MNRAS.465.1621P}, which gives a bulge-to-stellar mass fraction of $10\sim30\%$. Thus, our bulge fraction assumption of $13.1\%$ is consistent with observational constraints.  The bulge size choice follows from previous self-interacting dark matter simulations~\citep{2019MNRAS.490.2117R} that also implemented embedded disks. We note that this scale radius is consistent with the shorter axes of the MW bulge fitted using triaxial models~\citep{2002A&A...384..112L,2012A&A...538A.106R}, but larger than the $\sim 0.1$ kpc bulge radius fitted with spherical models~\citep{2013JCAP...07..016N,2014MNRAS.445.3133P}. 

At every snapshot, we assume $60\%$, $27\%$, and  $13\%$ of the total UM-predicted {\em stellar mass} to be distributed in the stellar disk, gaseous disk, and spherical bulge, respectively. The gaseous disk here is merely a more extended disk component. The combined disk potential has a mass that is a conservative lower bound of the central galaxy baryon mass as there is, on average, another $15\%\sim 30\%$ of gas mass in true galactic disks. However, it is unclear if the effects of subhalo mass loss due to a true viscous gas disk can be well-approximated by an analytic potential without proper treatment of hydrodynamics. It has also been shown in previous embedded disk simulations~\citep{2017MNRAS.471.1709G} and confirmed by our findings (Fig.~\ref{fig:Nratio}) that a $\gtrsim 30\%$ change in disk mass does not create sizable differences in subhalo abundance that overcome the host-to-host scatter. 

In our model, all characteristic sizes in the disk potential ($R_d, h_d, r_b$) evolve according to the mean mass-size relation evolution from 3D-HST~\citep{2014ApJ...788...28V}, $r\propto (1+z)^{-0.72}$ in their Table 2 for late-type galaxies with $M_{\ast}\sim 10^{10.75}\mathrm{M_{\astrosun}}$. Apart from the redshift evolution of the total disk mass and physical sizes, we assume that the three components' relative mass and size ratios in $\Phi(R, Z)$ are fixed throughout the simulation. We provide the detailed formulae of disk acceleration and density distribution in Appendix~\ref{sec:app:disk}.

To ensure that the embedded disk potentials are anchored to the MW host halo centers, we seed a `sink' particle~\citep{1995MNRAS.277..362B} at $z=3$ in every halo when the disks are initialized. We move the low-resolution particle ($m_{\rm DM, low}\sim 10^8\mathrm{M_{\astrosun}}$) with the closest distance to the host at $z=3$ (typical distance $\sim 3$ Mpc) to the halo center and initialize its velocity as the \textsc{Rockstar}-defined halo velocity. We note that the sink particle mass is comparable to the disk mass at initialization and becomes $<1\%$ of the disk mass at $z\lesssim1$. We modify \textsc{Gadget-2} to self-consistently implement the additional gravity from the disk potential on the dark matter particles. At every time step, the sink particle location is first used to initialize the center of the disk potential. Then, the gravitational force from the disk potential is applied to every particle other than the sink, in addition to the N-body forces in between those particles. Finally, the net gravitational force from all other particles (N-body forces and back reactions on the disk) is exerted on the sink particle to advance its velocity and position. The large particle mass of the sink particle combined with the equivalent mass of the disk potential (already $\gtrsim 10^3$ than high-resolution particles at $z=3$, Fig.~\ref{fig:sfh}) will keep the sink particle well-anchored to the MW center due to large dynamical friction~(see Fig.~\ref{fig:traj}). This procedure ensures a physical and stable implementation of the disk potential while adding negligible computational costs to the N-body simulation.

\begin{figure}
    \centering
	\includegraphics[width=\columnwidth]{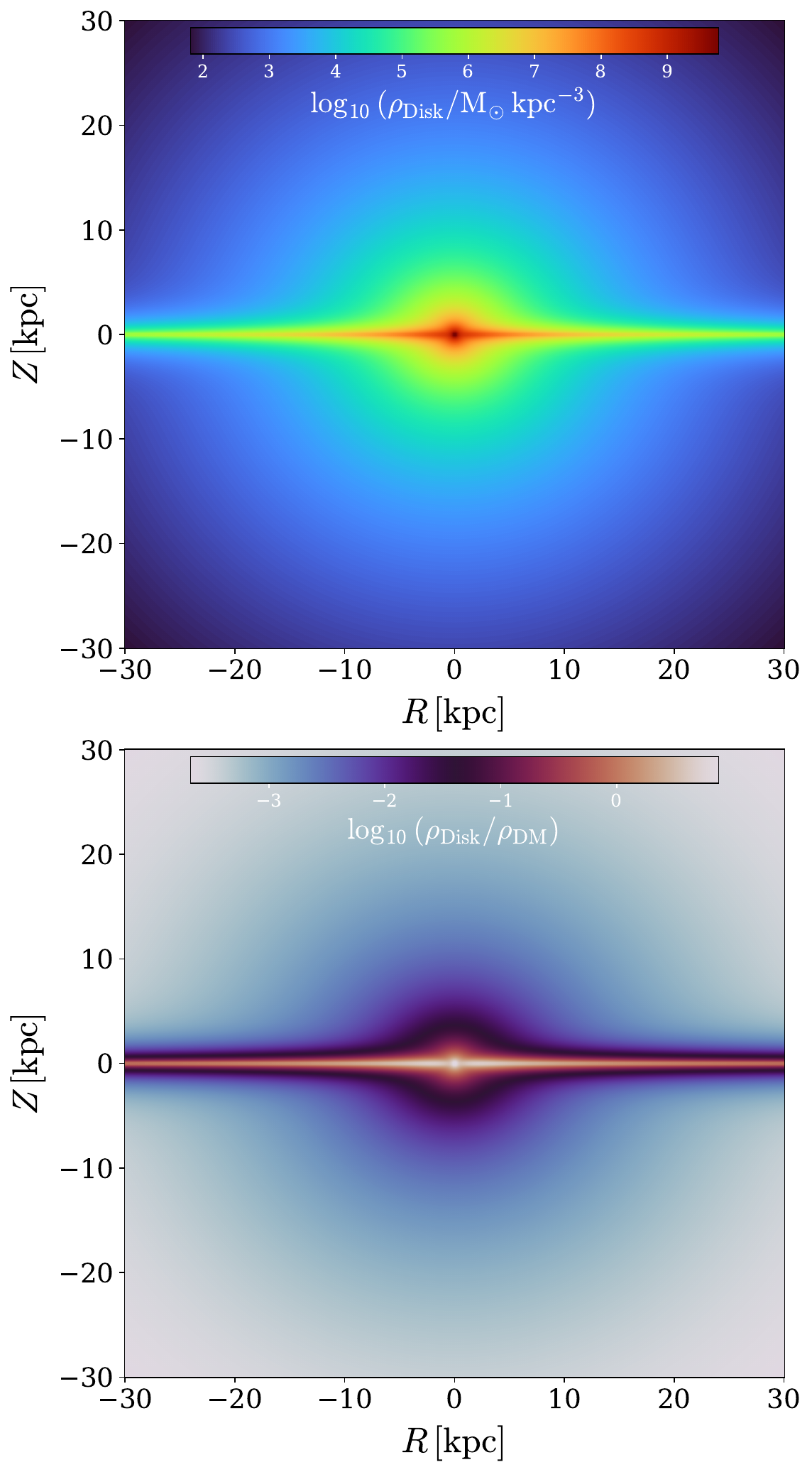}
    \caption{{\it Top panel:} Disk density map in the central 30 kpc region. We assume a total equivalent stellar mass of $2\times 10^{10}\mathrm{M_{\astrosun}}$. {\it Bottom panel:} Disk density to dark matter density ratio map. We assume the same disk mass as the top panel; we assume the NFW dark matter profile with $M_{200} = 9.7\times 10^{11}\mathrm{M_{\astrosun}}$ and $c_{200} = 9.37$ following MW observations~\citep{2020MNRAS.494.4291C}. }
    \label{fig:density}
\end{figure}

Using the disk potential, we can derive the corresponding density distribution from the analytic disk potential using the Poisson equation. We show an example 2D density distribution cross section in Fig.~\ref{fig:density}. We assume a total disk mass of $M_{\ast} = 2\times10^{10}\mathrm{M_{\astrosun}}$; the total disk density is shown in the top panel. In the bottom panel, we show the density ratio between the disk density and the underlying dark matter density to highlight where the disk dominates. For the dark matter density profile, we assume the best-fit Navarro-Frenk-White~\citep[NFW, ][]{1996ApJ...462..563N} density profile parameters of the MW from \citet{2020MNRAS.494.4291C}, with $M_{200} = 9.7\times 10^{11}\mathrm{M_{\astrosun}}$ and $c_{200} = 9.37$ (equivalent $R_{200} = 204$ kpc, $r_s = 22$ kpc) after considering adiabatic contraction due to the baryon disk.

\subsection{High-disk-mass subsample}
\label{sec:method:highmass}

To explicitly isolate the effect of the disk mass, we randomly choose nine halos out of our suite of 45 EDEN halos and re-simulate them with $\times 2.5$ higher disk masses (constant multiplicative factor) while fixing the shape of their disk growth histories. By doing this, we not only broaden our halo sample at high disk masses similar to our Galaxy, but we can also isolate the effect of varying disk masses on subhalo abundance. We show in Fig.~\ref{fig:sfh} the median stellar mass growth history of the high-disk-mass subsample. They are more consistent with recent MW observations (top panel) and previous embedded disk simulations (bottom panel) than the full EDEN suite. 

In the following, we refer to the set of 45 EDEN halos with UM-predicted $M_{\ast}(z)$ as `EDEN fiducial' (or EDEN for short) and the nine high-disk-mass subsample as `EDEN $M_{\ast, \rm Disk} \times 2.5$'; we refer to the DMO counterparts of `EDEN fiducial', i.e. the 45 SymphonyMilkyWay halos, as `Symphony'; we refer to the nine DMO counterparts of `EDEN $M_{\ast, \rm Disk} \times 2.5$' as `Symphony $M_{\ast, \rm Disk} \times 2.5$'.  

\section{The alluring myth of ``disruption''}
\label{sec:discussion:disruption}

Before presenting our results, we discuss some important issues related to their interpretation. The basic question is whether a reduction in subhalo peak (or present-day) mass functions should be interpreted as physical disruption---i.e., a reduction in subhalo abundances that would persist at any resolution level. Based on the recent literature, we argue that the safest and likely most accurate interpretation is instead that the disk enhances subhalo mass loss rates, such that objects are stripped below our resolution limit. 

\subsection{Background}
\label{sec:background}

The language that the field uses to describe mass loss in subhalos paints a picture in which subhalos are completely destroyed in a cataclysmic, discrete event: subhalos are said to ``disrupt.'' In previous embedded disk simulations~\citep{2017MNRAS.471.1709G,2019MNRAS.487.4409K}, the abundance suppression in the subhalo peak mass function (SPMF) is used as a diagnostic for `disruption' of subhalos instead of enhanced mass loss. The peak mass of subhalos does not change after infall, so if the subhalos only experience enhanced stripping and could still be tracked by the halo finder, there would be no suppression in the SPMF in the ideal case; whereas `disruption' of subhalos would lead to objects no longer detectable and cause a reduction in the SPMF. However, this picture is not correct for DMO CDM subhalos, and the concept of disruption can conflate three distinct scenarios. Colloquially, a subhalo ``disrupting'' can mean any of the following:

\begin{enumerate}
\item the point at which the subhalo is no longer self-gravitationally bound;
\item the point at which the subhalo finding tools can no longer identify the subhalo; or
\item the point at which the satellite galaxy the subhalo hosts has lost so much stellar mass that it would no longer be observable.
\end{enumerate}

At first blush, meanings 1 and 2 may seem reasonable or even synonymous. Early CDM simulations and analytic arguments seemed to favor a picture in which subhalos could truly disrupt (e.g.~\citealp{1994ApJ...433L..61G,2003ApJ...584..541H}). These arguments generally rely on comparing the characteristic energy injected into the subhalo by tidal shocks at pericenter to the total binding energy of the subhalo or by considering the properties of an isolated halo profile that was suddenly truncated. Although in detail these arguments break down \citep{2018MNRAS.474.3043V}, even in modern simulations, individual subhalos can suddenly disappear from subhalo catalogs at high resolutions or can be seen to experience runaway mass loss while they are still resolved by the subhalo finder (e.g., Fig.~16 and Fig.~5 in \citealt{2024ApJ...970..178M}, respectively). But this apparent ``disruption'' is illusory for most subhalos. While it is true that high-mass subhalos experience so much dynamical friction that they sink to the centers of their host within a few orbits (often remaining self-bound even after the merger finishes: \citealt{2016MNRAS.457.1208H,2024MNRAS.533.3811D,2024ApJ...970..178M}), lower mass subhalos experience much less dynamical friction~\citep[e.g.,][]{2016MNRAS.455..158V}. Idealized, high-resolution simulations show that low-mass subhalos with the high-internal-density ``cuspy'' profiles favored by CDM instead stay bound for much longer than the lifetime of the universe \citep{2010MNRAS.406.1290P,2018MNRAS.475.4066V,2018MNRAS.474.3043V,2020MNRAS.491.4591E,2021MNRAS.505...18E}. This is true even for simulations that include embedded disk potentials~\citep{2022MNRAS.509.2624G}. The loss of high-resolution subhalos from simulation catalogs can be explicitly shown to be a combination of subhalo finder failures and numerical non-convergence \citep{2024ApJ...970..178M}, while being highly dependent on resolution and subhalo finder.

The near limitless durability of DMO subhalos in $\Lambda$CDM is caused by their density profiles; models that lower the inner density --- for example, by modifying the nature of the dark matter particle --- can accelerate mass loss rates and even lead to true, resolved unbinding \citep{2021ApJ...920L..11N,2023MNRAS.519..384E,2023ApJ...949...67Y,2023ApJ...958L..39N,2024PhRvD.110b3019D}. This distinction makes it even more important to be careful about the term ``disruption'' in CDM contexts, as the term is a correct description of a physical process that is a hallmark of some alternative dark matter models. That said, even for non-$\Lambda$CDM models, it is worth being careful about what is and is not true disruption. The same processes that can numerically destroy DMO $\Lambda$CDM subhalos can accelerate the destruction of subhalos in other frameworks.

The third meaning above is on somewhat better theoretical footing for CDM subhalos: satellite galaxies can always lose enough mass to pass below a given optical observation limit. Studies of galaxy mass loss in hydrodynamic simulations suggest that for a given subhalo, there is a characteristic $M_{\rm vir}/M_{\rm peak}$ scale at which the subhalo's tidal radius begins to encroach on the satellite's stars and thus its stellar mass loss rate suddenly accelerates~\citep{2016ApJ...833..109S}. But even this meaning has some conceptual and logistical issues. Conceptually, the existence of a characteristic scale in the mass-loss rate does not necessarily imply disruption, just that there is essentially no stellar mass loss soon after infall. Logistically, adopting a star-based approach means that one needs to have a model for the energy/radial distribution of stars in a subhalo, as the end state of a subhalo's satellite galaxy varies drastically based on these quantities \citep[e.g.][]{2024ApJ...968...89E}. Additionally, numerical problems associated with the setup of most hydrodynamic simulations mean that they resolve the mass loss of satellite galaxies far more poorly than the disruption of subhalos \citep{2019MNRAS.488L.123L,2020MNRAS.493.2926L,2021MNRAS.508.5114L,2023MNRAS.525.5614L}. This is not an intractable problem, but it does complicate the issue to the point where it is out of scope for this work.

\subsection{Interpreting EDEN Results}

Thus, when evaluating the impact of the disk potentials on subhalo mass functions in our simulations, we do not say that decreased subhalo mass function amplitudes imply that the disk completely disrupts subhalos. Instead, we interpret this as evidence that disks accelerate subhalos' mass loss rates.

This distinction holds even when we compare the subhalo \emph{peak} mass function between DMO and Disk runs (Section~\ref{sec:results:SPMF}). The peak halo mass $M_{\rm peak}$ is the peak historical value of the halo virial mass $M_{\rm vir}$, which reflects the pre-infall mass of subhalos before experiencing significant tidal stripping from their hosts. In the case of a suppressed SPMF, the increased mass loss rate causes subhalos to approach the numerical-driven limits of the sample more quickly. Given the discussion in Section~\ref{sec:background}, one may question the wisdom in evaluating SPMF at all: after all, at low masses, the SPMF is just the infall mass function minus subhalos that have either numerically disrupted or have been lost by the subhalo finder. Skepticism is warranted, but $M_{\rm peak}$-based measurements serve a legitimate purpose. The fact that subhalos must lose large amounts of dark matter mass prior to losing any stellar mass means that infall- and peak-based mass definitions tend to be better predictors of galaxy properties than the current halo mass~\citep[e.g.,][]{2013ApJ...771...30R}. In particular, $M_{\rm peak}$-selected subhalo samples have much higher small-radius number densities around their hosts than $M_{\rm vir}$-selected samples (see review in Section 5.2 of \citealp{2024ApJ...970..178M}), in line with observations. 

Indeed, even when analyzed with \textsc{Symfind}, which is capable of tracing $\approx 90\%$ of the subhalos at our minimum mass threshold to the mass scales at which galaxy disruption is likely to begin \citep{2024ApJ...970..178M}, the median value of $M_{\rm vir}/M_{\rm peak}$ in our sample is approximately 0.2 in both SymphonyMilkyWay and EDEN. This $M_{\rm vir}/M_{\rm peak}$ threshold is a factor of $2$ to $4$ higher than the masses that hydrodynamic simulations~\citep{2016ApJ...833..109S} and empirical models~\citep{2018MNRAS.477.1822M,2019MNRAS.488.3143B,2024ApJ...976..119W} predict that stellar mass loss should occur. Thus, the vast majority of subhalos considered in this paper would likely still host visible galaxies (if they had one to begin with). This means that the differences in $M_{\rm peak}$ functions discussed below would likely qualitatively extend to stellar mass-selected samples, even if the quantitative translation would require significant modeling work (see Section~\ref{discussion:interpretting} for further discussion).

\begin{figure*}
    \centering
	\includegraphics[width=1.35\columnwidth]{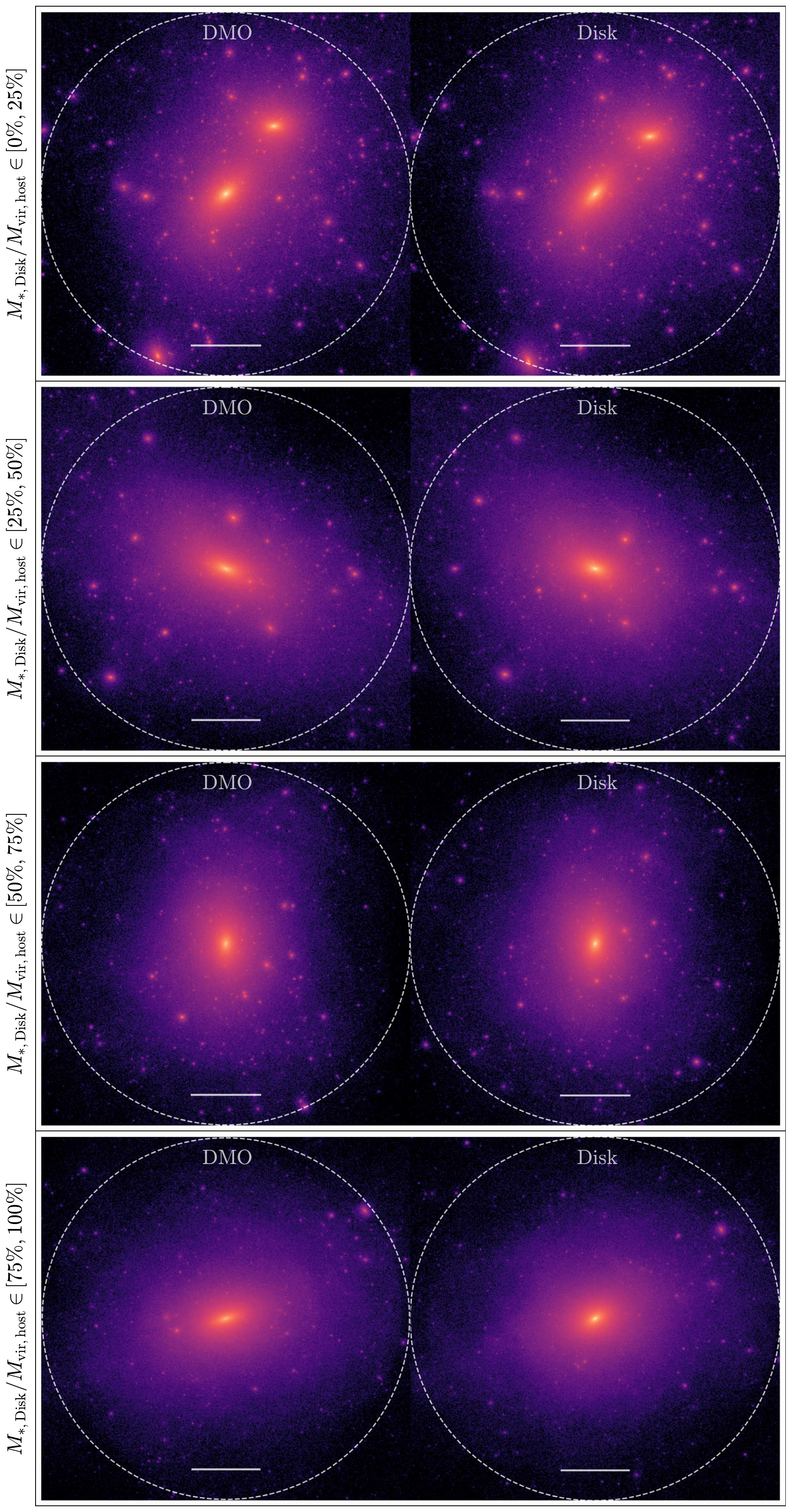}
    \caption{Dark matter density maps for four hosts in four quartiles of $M_{\ast, \rm Disk}/M_{\rm vir, host}$ increasing from top to bottom. The left column shows the SymphonyMilkyWay DMO halos and the right column shows their embedded disk counterparts in EDEN. The DMO and Disk density maps for each halo share the same color scales. The dashed circles mark out the virial radii and the scale bars denote $50\ \mathrm{kpc}\,h^{-1}$. It is apparent that subhalo abundance suppression becomes stronger with increasing $M_{\ast, \rm Disk}/M_{\rm vir, host}$, especially in the inner regions of each host. The central density of the hosts also increases more from top to bottom due to higher adiabatic contraction with larger embedded disk potentials.}
    \label{fig:map}
\end{figure*}

\begin{figure*}
    \centering
	\includegraphics[width=1.6\columnwidth]{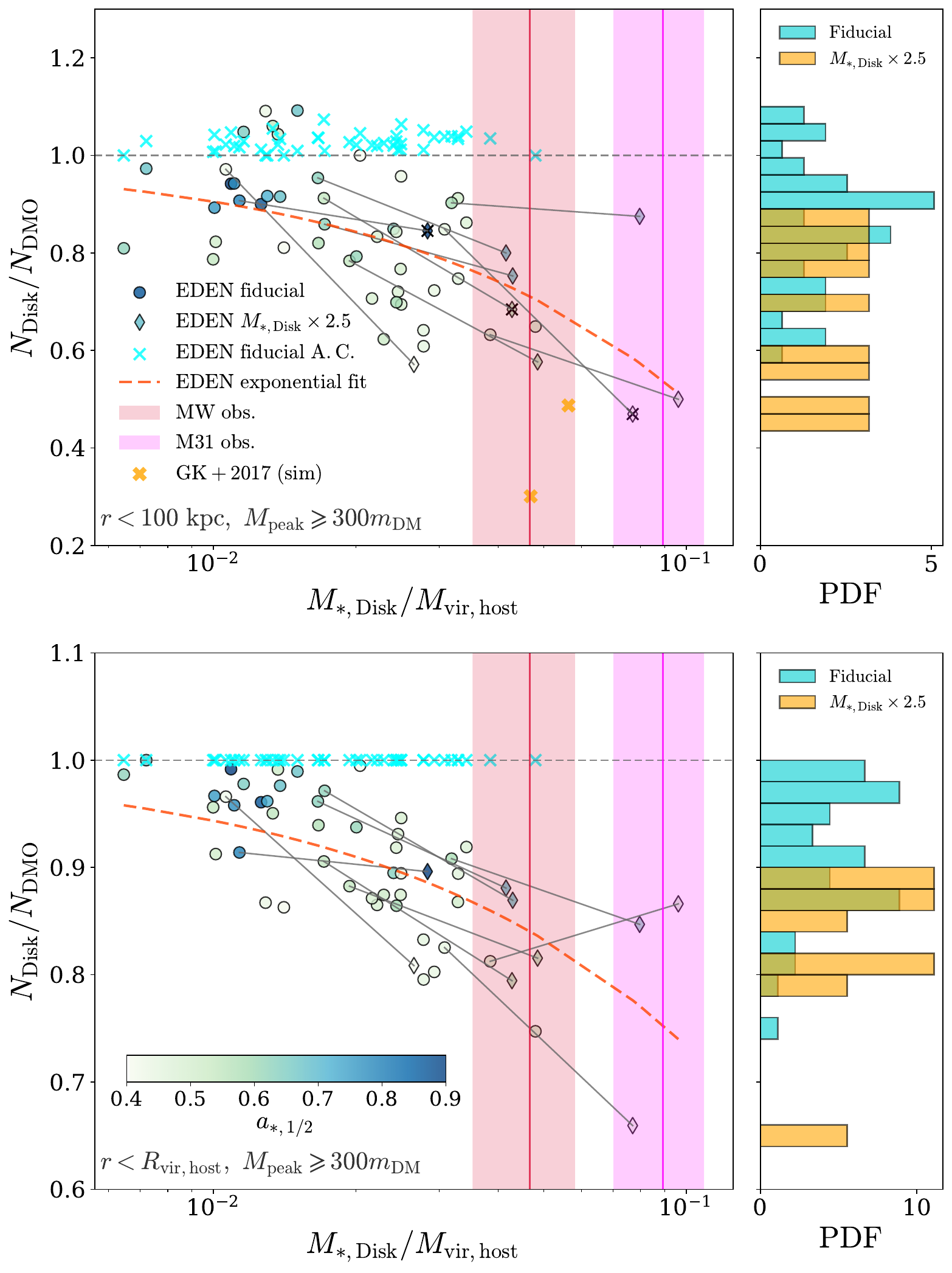}
    \caption{Subhalo count ratios between EDEN Disk and Symphony DMO simulations for resolved subhalos with $M_{\rm peak}\geqslant 300 m_{\rm DM}$ ($1.2\times 10^{8}\mathrm{M_{\astrosun}}$). The top panel shows only subhalos within 100 kpc of their hosts, while the bottom panel shows all subhalos within the virial radii of their hosts. Each colored circle is a pair of Symphony/EDEN hosts, with $M_{\ast, \rm Disk}/M_{\rm vir, host}$ values from Symphony. The nine MW-mass hosts we re-simulated with $M_{\ast, \rm Disk} \times 2.5$ are shown with diamonds and lines connecting them with their EDEN fiducial runs. The color bar indicates the stellar half-mass formation scale factor. The Spearman correlation coefficient between $N_{\rm Disk}/N_{\rm DMO}$ and $\log_{10}(M_{\ast, \rm Disk}/M_{\rm vir, host})$ for EDEN fiducial is $r_S = -0.60$ for $r<100$ kpc and $r_S = -0.74$ for $r<R_{\rm vir, host}$, which are statistically significant negative correlations. We also provide exponential fits to these relations whose best-fit parameters are shown in Table~\ref{tab:fit}. The Spearman $r_S$ and exponential fit are carried out on the combined set of EDEN fiducial and EDEN $M_{\ast, \rm Disk}\times 2.5$ simulations. The cyan crosses are the expected number ratio of subhalos \emph{in the absence of disk-induced mass loss} due to adiabatic contraction (A.C.) from the disk potential. The vertical red (MW, \citealt{2020MNRAS.494.4291C}) and magenta (M31, \citealt{2010MNRAS.406..264W,2012A&A...546A...4T}) bands mark the observational constraints for MW/M31. Orange crosses in the top panel denote the two DMO/Embedded-disk halo pairs from \citet{2017MNRAS.471.1709G} whose hydrodynamic counterparts were simulated with FIRE-2~\citep{2023ApJS..265...44W}. Disk stripping dominates over adiabatic contraction across the range of $M_{\ast, \rm Disk}/M_{\rm vir, host}$ covered by EDEN.}
    \label{fig:Nratio}
\end{figure*}

\section{Results}
\label{sec:results}

Here, we present the key results of EDEN. As we have reviewed in Section \ref{sec:discussion:disruption}, there are important caveats regarding whether the reduction in subhalo mass functions in our disk simulations represents physical ``disruption.'' Due to these complications, we explicitly avoid the term ``disruption'' and instead describe how the disk suppresses subhalo peak and present-day mass functions; we also study the spatial and orbital dependence of these effects. In Section~\ref{sec:results:mdisk}, we present the magnitude of subhalo mass function suppression due to the disk as a function of host stellar-to-halo mass ratio and demonstrate that a larger disk creates more subhalo mass loss; in Section~\ref{sec:results:SPMF} we present the subhalo peak mass functions; in Section~\ref{sec:results:radial} we present the subhalo radial distance and pericentric distance distributions; in Section~\ref{sec:results:tinfall} we present the subhalo infall time distributions.

\subsection{Disk-to-halo mass ratio as a critical factor for subhalo abundance}
\label{sec:results:mdisk}

To intuitively visualize the effect of different disk-to-halo mass ratios on subhalo abundance, we show in Fig.~\ref{fig:map} the projected dark matter density maps for four example halos from the EDEN and Symphony comparing their DMO and Disk counterparts. The four example halos fall in four quartiles of disk-to-halo mass ratios; this range creates a visually significant change in subhalo abundance. The larger the disk-to-halo mass ratios, the stronger the suppression of substructures, especially in the central regions of the host. We also note that the host central density becomes higher, and their shapes become rounder due to adiabatic contraction. The rounder halo shapes are qualitatively consistent with results from previous embedded disk simulations~\citep{2017MNRAS.471.1709G,2019MNRAS.487.4409K}; we show the axis ratios in Appendix~\ref{sec:app:axis}.

To further quantify the effects of disk-to-halo mass ratio in subhalo abundance,  Fig.~\ref{fig:Nratio} shows the subhalo count ratios within 100 kpc and $R_{\rm vir, host}$ between the EDEN fiducial Disk runs and their corresponding Symphony DMO runs (45). In both radial ranges, there are clear trends that subhalo abundance suppression is more effective in halos with larger $M_{\ast, \rm Disk}/M_{\rm vir, host}$. Since the halo mass range is quite narrow for the 45 SymphonyMilkyWay hosts (Fig.~\ref{fig:smhm}), the primary effect is of changing the disk mass; we have checked that we get almost identical results if we instead used $M_{\ast, \rm Disk}$ (see also similar trends in hydrodynamic simulation FIRE-2, Fig. 6 in ~\citealt{2020MNRAS.491.1471S}). 

At low disk mass, $M_{\ast, \rm Disk}/M_{\rm vir, host}\lesssim 0.02$, most EDEN halos demonstrate $\lesssim 20\%$ fewer subhalos than their disk-less counterparts within 100 kpc and $\lesssim 10\%$ within $R_{\rm vir, host}$. There are a few halos with $N_{\rm Disk}/N_{\rm DMO}>1$. Physically, the addition of the disk can cause the adiabatic contraction of dark matter~\citep{1986ApJ...301...27B,1987ApJ...318...15R,2004ApJ...616...16G} and draw more subhalos into the host. Because the disk has two effects, one which increases subhalo abundance and another that decreases it, it is possible that if a low-mass disk's effect on subhalo mass loss is weak, its host might end up having $N_{\rm Disk}/N_{\rm DMO}>1$.  

We apply the \citet{1986ApJ...301...27B} adiabatic contraction model to the Symphony DMO hosts to quantify the expected enhancement in subhalo counts due to adiabatic contraction. This abundance represents the expected number of subhalos within 100 kpc or $R_{\rm vir, host}$ after including the disk as if the disk potential only contracts subhalo orbits and does not increase their mass loss rates. Assuming spherical symmetry, circular subhalo orbits, and conservation of angular momentum, we have \footnote{The disk mass profile $M_{\rm Disk}(<r)$ for each halo is derived by integrating the axisymmetric disk density $\rho (R,Z)$ (Fig.~\ref{fig:density}, each host with their own UM-predicted total $M_{\ast}$) radially where $r = \sqrt{R^2 + Z^2}$.}
\begin{equation}
    \label{eq:AC}
    \left(1 + \frac{M_{\ast, \rm Disk}}{M_{\rm vir, host}}\right) M_{\rm DM}(<r_0) r_0 = \left[M_{\rm DM}(<r_0)+M_{\rm Disk}(<r_f)\right] r_f.
\end{equation}
Here $r_0$ is the distance to the MW host center in the DMO run, $M_{\rm DM}(<r_0)$ is the enclosed dark matter mass within $r_0$, which is adiabatically contracted to a smaller sphere with radius $r_f$ after adding the disk potential. The mass component of the disk potential before contraction is assumed to follow that of the dark matter in the DMO run. We assume that the locations of the subhalos contract in the same manner as the underlying dark matter density profile of their host, such that subhalos located at $r_0$ in SymphonyMilkyWay are expected to be at $r_f$. We show the contracted subhalo count ratios within 100 kpc and $R_{\rm vir, host}$ in Fig.~\ref{fig:Nratio}, which creates a $\sim 5\%$ subhalo count boost within 100 kpc and no effect out to $R_{\rm vir, host}$. The EDEN hosts having $N_{\rm Disk}/N_{\rm DMO} > 1$ within 100 kpc all have low disk-to-halo mass ratios and their subhalo abundance increase is consistent with the predictions of adiabatic contraction.

At the massive disk mass end, where $M_{\ast, \rm Disk}/M_{\rm vir, host}\gtrsim 0.02$, $N_{\rm Disk}/N_{\rm DMO}\ll1$ and disk stripping effects dominate over adiabatic contraction. The two EDEN halos with heavy disks that fall within the observational errors of the MW (blue band, \citealt{2020MNRAS.494.4291C}), have $30\sim 40\%$ fewer subhalos within 100 kpc and $20\sim 30\%$ fewer subhalos within $R_{\rm vir, host}$. We also show the subhalo count ratios for the nine halos in EDEN $M_{\ast,\rm Disk}\times 2.5$ and connect them with EDEN fiducial values. Three of the $M_{\ast, \rm Disk} \times 2.5$ hosts have $M_{\ast, \rm Disk}/M_{\rm vir, host}$ within the observational uncertainties of M31~\citep{2010MNRAS.406..264W,2012A&A...546A...4T} and show even larger suppression on average. The average subhalo abundance suppression in the nine halos with heavier disks is more prominent in both radial ranges (side histograms), all having lower $N_{\rm Disk}/N_{\rm DMO}$ with increasing $M_{\ast, \rm Disk}/M_{\rm vir, host}$. 

Three of these EDEN $M_{\ast,\rm Disk}\times 2.5$ hosts fall within the M31 disk-to-halo mass ratio range and two of them produce $\gtrsim 50\%$ subhalo abundance suppression within 100 kpc. The one outlier (Halo 675) that shows a slight decrease in subhalo counts at $r<100$ kpc but an increase in count at $r<R_{\rm vir, host}$ after inserting a $M_{\ast,\rm Disk}\times 2.5$ disk is undergoing a major merger and experiences a boost in outskirt subhalo counts due to its infalling companion. 

Apart from the effect of varying disk-to-halo mass ratio on subhalo abundance, a secondary trend with disk half-mass formation scale $a_{\ast, 1/2}$ is also present in the color coding of the data points in Fig.~\ref{fig:Nratio}. This is a consequence of heavier disks also growing their disks earlier (Fig.~\ref{fig:smhm}), which could impact their subhalo populations longer and lead to more suppression as the disk preferably loses subhalos with earlier infall times (Fig.~\ref{fig:tinfall}). Therefore, the additional variations in disk formation history could further enhance the suppression in subhalo abundance caused by varying disk masses.

\begin{figure*}
    \centering
	\includegraphics[width=2\columnwidth]{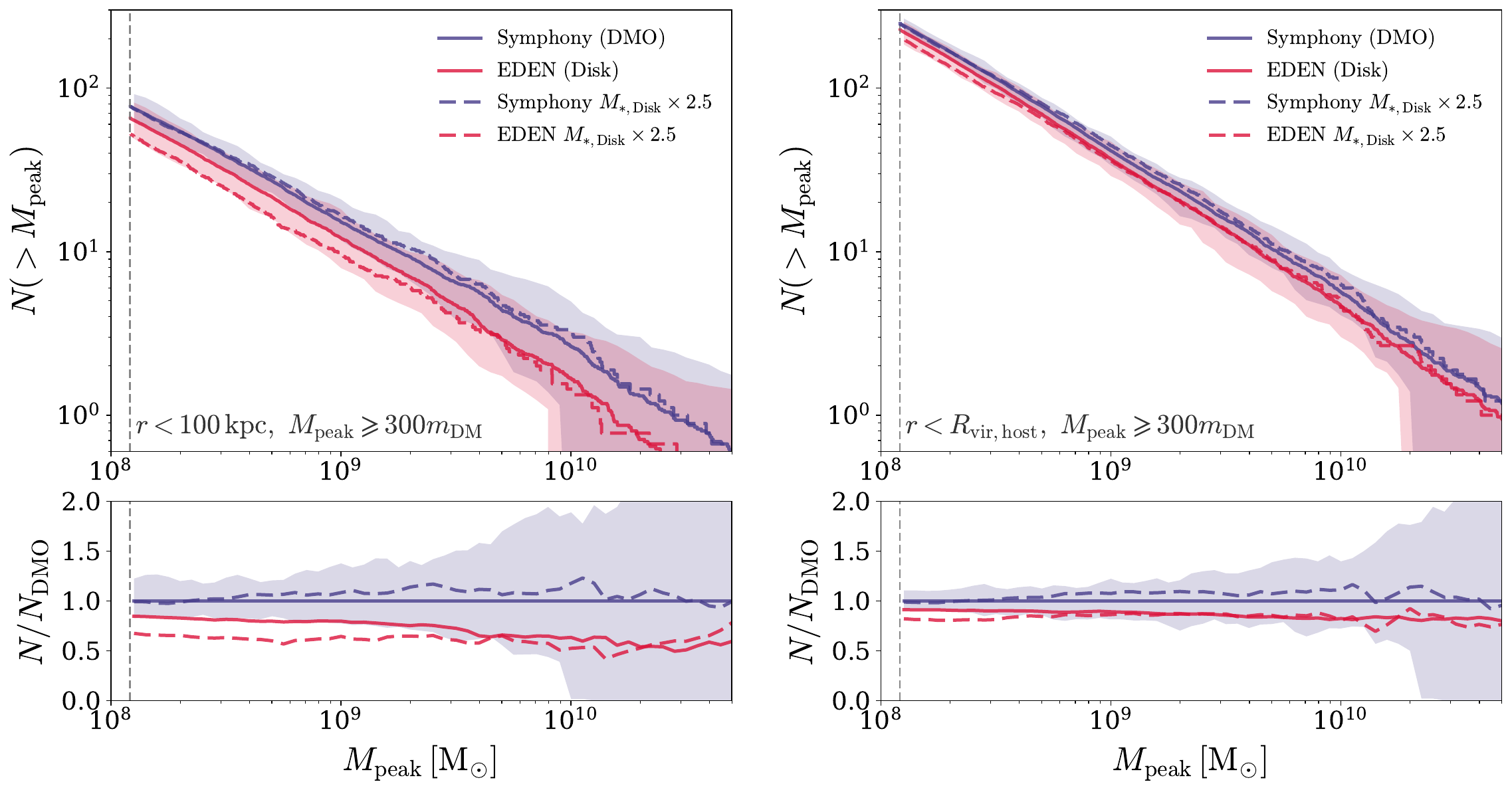}
    \caption{The subhalo peak mass function (SPMF) for $z=0$ surviving subhalos. Left and right panels show the SPMF functions within 100 kpc and $R_{\rm vir, host}$, respectively. The bottom sidebars in each panel show the SPMF ratios relative to the median DMO SPMF in that radial range. In each panel, the solid curve is the median SPMF for Symphony (blue, DMO) and EDEN fiducial (red, Disk). The shaded region in each panel shows the $68\%$ host-to-host scatter in Symphony or EDEN. The red dashed curves are the average SPMF for the nine hosts in EDEN $M_{\ast, \rm Disk}\times 2.5$, while the blue dashed curves are their DMO counterparts in Symphony. Subhalo abundance suppression is mass-independent only in the high-disk-mass subsample, and high-mass subhalos are preferentially lost in EDEN fiducial. The vertical dashed lines mark out the resolution limit of subhalos in this work ($M_{\rm peak}\geqslant 300 m_{\rm DM}\sim 1.2\times 10^{8}\mathrm{M_{\astrosun}}$).}
    \label{fig:SHMF}
\end{figure*}

To summarize our findings, disk effects on subhalo abundance are sensitive to the disk-to-halo mass ratio of the host. The MW and M31 are amongst the most efficient systems in subhalo abundance suppression due to their atypically large $M_{\ast, \rm Disk}/M_{\rm vir, host}$. Disk effects on subhalo abundances are weak for small $M_{\ast, \rm Disk}/M_{\rm vir, host}\lesssim 0.02$ hosts. For large $M_{\ast, \rm Disk}/M_{\rm vir, host}\gtrsim 0.02$ hosts, disk effects dominate over adiabatic contraction and become stronger with increasing disk-to-halo mass ratios. We fit the $\log_{10}(M_{\ast, \rm Disk}/M_{\rm vir, host})$-$N_{\rm Disk}/N_{\rm DMO}$ relation jointly for the 45 EDEN fiducial halos and nine EDEN $M_{\ast, \rm Disk}\times 2.5$ halos, using an exponential functional form whose best-fit parameters are provided in Table~\ref{tab:fit}. The Spearman rank correlation coefficients for these joint relations are $r_S = -0.60$ within 100 kpc and $r_S = -0.74$ within $R_{\rm vir, host}$, both being statistically significant negative correlations between subhalo count suppression and $M_{\ast, \rm Disk}/M_{\rm vir, host}$. 

\begin{table}
    \centering
    \begin{tabular}{lcc}
			\hline
            \hline
			Range & $a$ & $x_0$\\
			\hline
            $r<100$ kpc & $2.64\pm 1.05$ & $0.61\pm 0.10$ \\
            $r<R_{\rm vir, host}$ & $1.21\pm 0.39$ & $0.67\pm 0.10$ \\
			\hline
    \end{tabular}
    \caption{Best-fit parameters for $y = N_{\rm Disk}/N_{\rm DMO}$ as a function of $x = \log_{10}(M_{\ast, \rm Disk}/M_{\rm vir, host})$ shown in Fig.~\ref{fig:Nratio}. These fits are for the combined sample of EDEN fiducial (45) and EDEN $M_{\ast, \rm Disk}\times 2.5$ (9). The adopted functional form of the fit is $y = 1 -a\exp{(x/x_0)}$.}
    \label{tab:fit}
\end{table}

\subsection{Subhalo peak mass functions}
\label{sec:results:SPMF}

\begin{figure*}[!htbp]
    \centering
	\includegraphics[width=2\columnwidth]{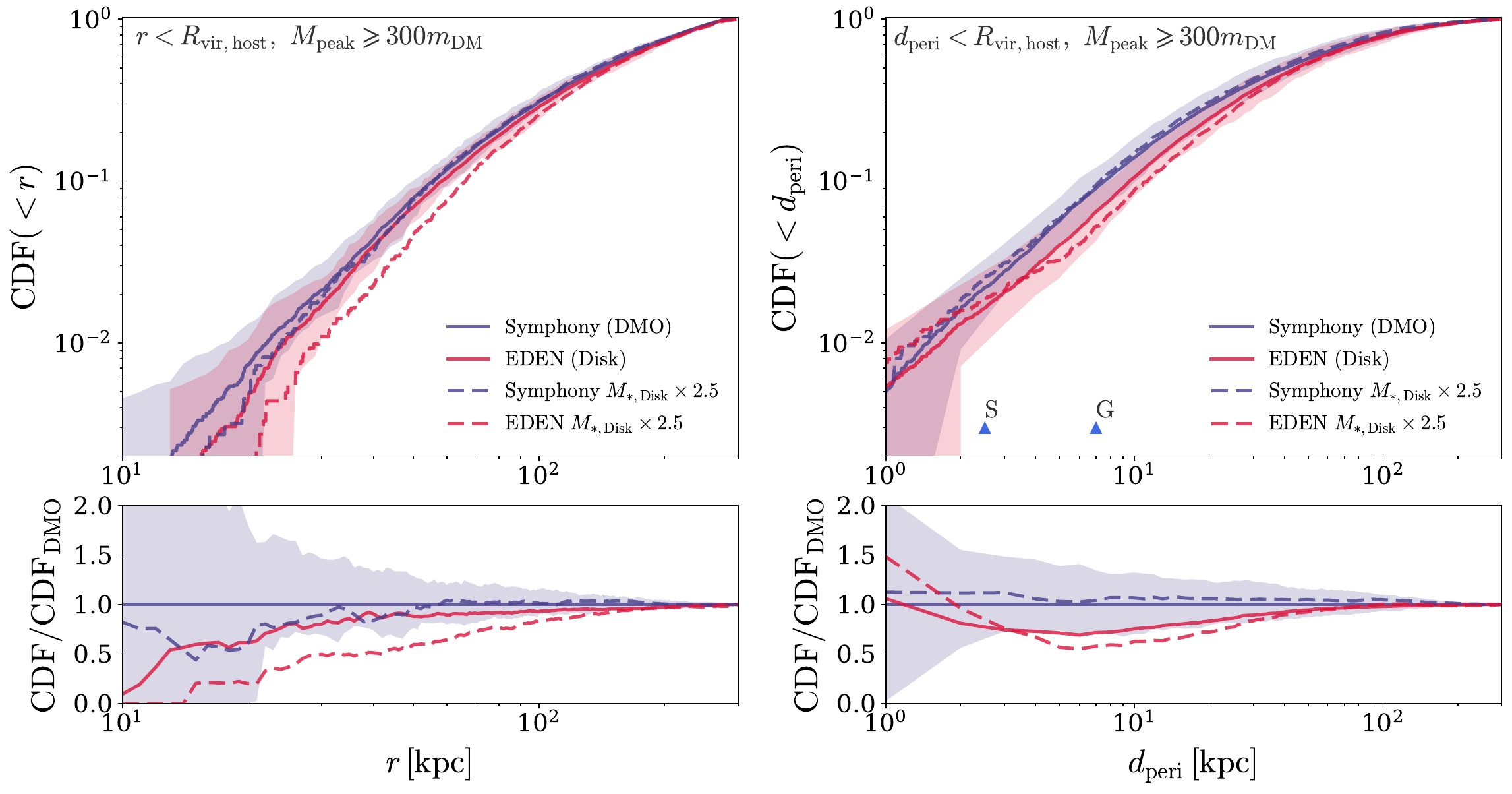}
    \caption{Subhalo radial cumulative distributions for resolved subhalos with $M_{\rm peak}\geqslant 300 m_{\rm DM}$ ($1.2\times 10^{8}\mathrm{M_{\astrosun}}$). The left panels show the subhalo abundance versus the $z=0$ distance ($r$) to the MW host center, and the right panels show the radial cumulative function of their orbit pericenter distances ($d_{\rm peri}$). Solid curves show average subhalo radial distributions for the Symphony (blue) and EDEN (red); shaded regions with the same color denote the corresponding $68\%$ host-to-host scatter. The dashed red curve shows the average subhalo radial distribution for the nine halos in EDEN $M_{\ast, \rm Disk}\times 2.5$, and the blue dashed curve shows the corresponding DMO subsample in Symphony. The lower side panels show the CDF ratio of each curve relative to Symphony. The triangles in the right panel denote the scale radii of the `stellar' (S, 2.5 kpc) and `gaseous' (G, 7 kpc) disks of the embedded potential. Subhalo abundance suppression becomes weaker for subhalos penetrating into the inner parts of the disk potential. We caution that these radial distributions may change significantly due to numerical resolution \citep{2024ApJ...970..178M}.}
    \label{fig:radial}
\end{figure*}

Following the total subhalo count ratios shown in Section~\ref{sec:results:mdisk}, we present the subhalo peak mass function (SPMF) for surviving subhalos within 100 kpc and $R_{\rm vir, host}$ in Fig.~\ref{fig:SHMF}. Comparing the median SPMF in Symphony and EDEN fiducial, 
the impact of the disk is stronger within 100 kpc than within  $R_{\rm vir, host}$, indicating radially dependent suppression (see Fig.~\ref{fig:radial}).
In addition to the change in amplitude, the slope of the SPMF is steeper in EDEN than in Symphony, indicating that the disk is more effective at removing mass from high-mass subhalos than it is for low-mass subhalos.

We also compare the median SPMF of the nine halos in the EDEN $M_{\ast, \rm Disk} \times 2.5$ subsample with their Symphony DMO counterparts. Their DMO counterparts, labeled as `Symphony $M_{\ast, \rm Disk} \times 2.5$', show a similar median SPMF to the full Symphony suite, suggesting that this nine-halo subsample is an unbiased representation of the 45 Symphony halos. However, the SPMF of the EDEN $M_{\ast, \rm Disk} \times 2.5$ halos is more suppressed than the EDEN fiducial median SPMF at all resolved subhalo masses for both $r<100$ kpc and $r<R_{\rm vir, host}$, which is another manifestation of subhalo abundance suppression being sensitive to the disk-to-halo mass ratio (Fig.~\ref{fig:Nratio}). 

Furthermore, the SPMF of EDEN $M_{\ast, \rm Disk} \times 2.5$ is independent of the subhalo masses, in line with previous embedded disk simulations that targeted MW-like heavy disk systems~\citep{2017MNRAS.471.1709G,2019MNRAS.487.4409K}. As we show in Appendix~\ref{sec:app:RF}, the Random Forest model N18~\citep{2018ApJ...859..129N}  that was trained on the two halos in \citet{2017MNRAS.471.1709G} with heavy disk masses shows similar levels of subhalo abundance suppression as EDEN $M_{\ast, \rm Disk} \times 2.5$ (using AHF halo catalogs). These lines of evidence are in contrast to the median SPMF of EDEN fiducial and indicate that the extra suppression in EDEN $M_{\ast, \rm Disk} \times 2.5$ mainly occurs in lower-mass subhalos ($M_{\rm peak}\lesssim 10^{10}\mathrm{M_{\astrosun}}$). These lower-mass subhalos would have otherwise survived in EDEN fiducial due to their weaker dynamical friction~\citep{1943ApJ....97..255C,1993MNRAS.262..627L} than higher-mass subhalos, sinking slower to the host center where the disk tidal field is strong.

\subsection{Subhalo radial distribution}
\label{sec:results:radial}

As shown in Section~\ref{sec:results:SPMF}, subhalo abundance suppression in our disk simulations is radially dependent, with most of the suppression happening at the inner 100 kpc. In this section, we show the subhalo radial distributions in terms of their final distances to their MW hosts' center as well as their orbital pericenter distances.

In the left panel of Fig.~\ref{fig:radial}, we show the cumulative radial distributions of subhalo distances to their hosts at $z=0$. The median EDEN fiducial CDF is more suppressed at all radii than the median CDF for Symphony, with slightly more suppression happening towards smaller radii ($r\lesssim 50$ kpc). However, the lowering of the CDF median is within the $68\%$ host-to-host scatter of the 45 halos. EDEN $M_{\ast, \rm Disk}\times 2.5$ magnifies the radial CDF suppression effect, with most of the subhalo abundance suppression happening at $r\lesssim 100$ kpc relative to their Symphony DMO counterparts. These findings indicate that the surviving subhalo population may not have significantly altered radial profiles due to the disk unless in very high disk-to-halo mass ratio systems like in the MW or EDEN $M_{\ast, \rm Disk}\times 2.5$.

In the right panel of Fig.~\ref{fig:radial}, we show the radial distributions for subhalo orbital pericenter distances. To derive pericenter distances, we use cubic splines to interpolate between the lookback time of simulation snapshots and the location of subhalos relative to their hosts. The interpolation is required since the snapshot cadence ($\sim 160$ Myr at $z=0$ and $\sim 40$ Myr at $z=3$) is usually too coarse to resolve the time period near each subhalo's pericenter.  The interpolated subhalo and host halo trajectories were then evaluated on 10 Myr cadences during the period when both the subhalo and their host halo existed. The minimum distance between the subhalo and host halo on this 10 Myr grid is then defined as the pericenter distance. 

From Fig.~\ref{fig:radial}, we can see most of the abundance suppression happens for subhalos with $d_{\rm peri}\lesssim100$ kpc. The pericenter distance CDF for EDEN fiducial has a larger suppression than in its radial distance CDF (left panel) and is $\sim 1\sigma$ lower than the Symphony pericenter CDF. This reveals that the disk potential preferentially removes mass from subhalos that came closer to it during pericenter, whereas subsequent orbital motion smears out this signal in the radial CDF. Increasing disk-to-halo mass ratios creates an even larger suppression at fixed $r$ as shown in the EDEN $M_{\ast, \rm Disk}\times 2.5$ CDF, with most of the additional suppression happening at $d_{\mathrm {peri}} \lesssim 50$ kpc.

\begin{figure*}
    \centering
	\includegraphics[width=2\columnwidth]{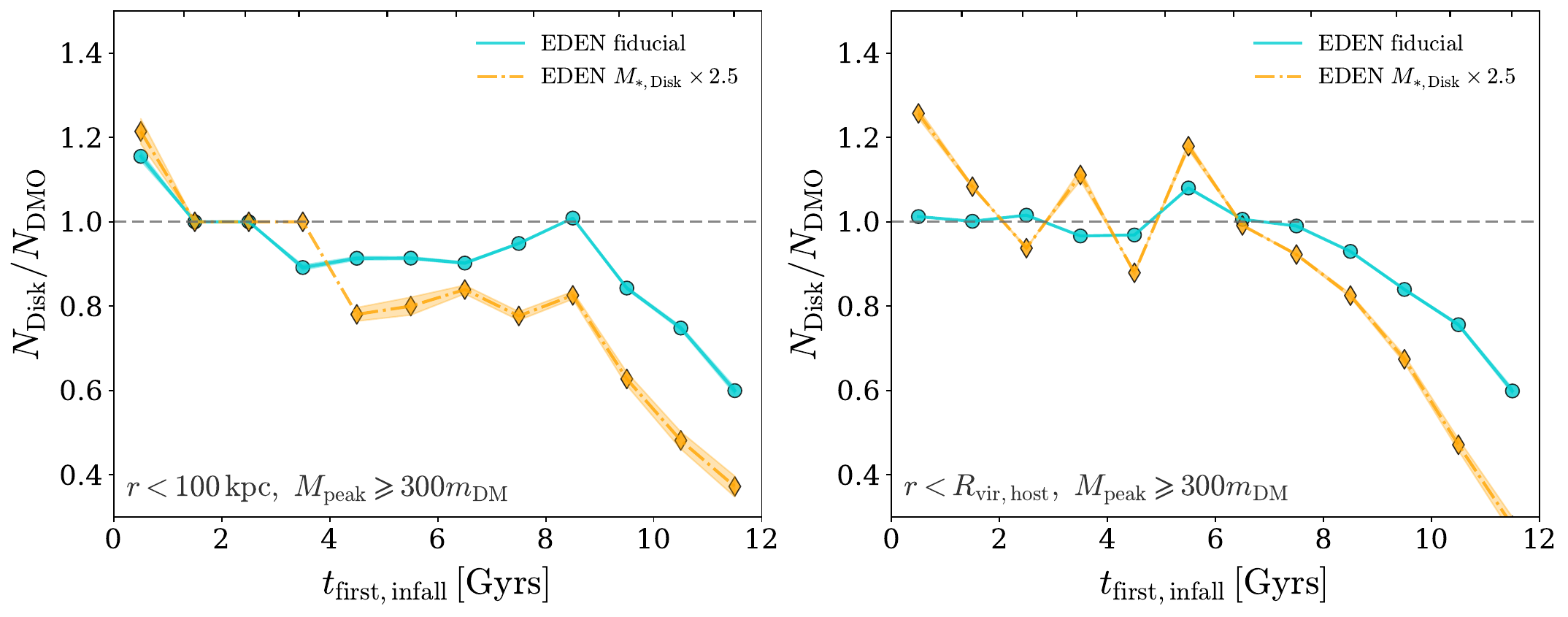}
    \caption{Subhalo count ratio as a function of their first-infall lookback-time into the MW host for resolved subhalos with $M_{\rm peak}\geqslant 300\,m_{\rm DM}$ ($1.2\times 10^{8}\mathrm{M_{\astrosun}}$). The left panel shows subhalos within 100 kpc, and the right panel shows subhalos within $R_{\rm vir, host}$. The orange circles denote the EDEN fiducial halos, and the turquoise diamonds denote the nine halos in EDEN $M_{\ast, \rm Disk}\times 2.5$. The shaded bands are Poisson errors on the subhalo count ratios. The subhalos removed from our sample by the inclusion of a disk are preferentially those with earlier infall times.}
    \label{fig:tinfall}
\end{figure*}

To summarize, halos with disks have fewer subhalos that fall on orbits with pericenter distances $d_{\rm peri}\lesssim100$ kpc. Increasing the disk mass increases subhalo abundance suppression strength, especially at $d_{\rm peri}\lesssim 50$ kpc. We caution that these radial trends in subhalo abundances should be interpreted as qualitative instead of quantitative as subhalo radial distributions can change significantly~\citep{2024ApJ...970..178M}.

\subsection{Subhalo infall times}
\label{sec:results:tinfall}

As we have shown in the previous sections, subhalo abundance suppression in the presence of a galactic disk is sensitive to the disk-to-halo mass ratio, the subhalo peak mass, and subhalo orbit pericenter distances. As found in Phat ELVIS~\citep{2019MNRAS.487.4409K}, subhalos that fell in earlier to their hosts are more preferentially stripped than later-infalling subhalos (see also Fig. 1 in \citealt{2017MNRAS.472..657J}). The physical intuition behind this effect is that earlier-infalling subhalos fell in when their host halos were smaller and have systematically smaller pericenters~\citep{2010MNRAS.403.1072W}, leading to stronger tidal stripping from the disk that enhances their mass loss. In the following, we investigate if this effect is generic in EDEN as Phat ELVIS used a MW-like heavy disk mass (Fig.~\ref{fig:smhm}).

In Fig.~\ref{fig:tinfall}, we show the subhalo count ratio in EDEN versus Symphony as a function of the lookback time of the first-infall into their MW-mass hosts ($r<100$ kpc in the left panel and $r<R_{\rm vir, host}$ in the right panel). In both radial ranges, subhalo abundance suppression mainly happens for early-infalling subhalos with $t_{\rm first, infall} > 8$ Gyrs, consistent with the findings in Phat ELVIS~\citep{2019MNRAS.487.4409K}. There are about $\sim 10\%$ of subhalos with $r<100$ kpc and fell in from $4\sim8$ Gyrs ago that are lost from disk tidal stripping. However, this suppression effect is washed out if we consider all subhalos within $R_{\rm vir, host}$ that fell in $4\sim 8$ Gyrs ago. The nine halos in EDEN $M_{\ast, \rm Disk}\times 2.5$ further enhance the infall time distribution patterns of EDEN fiducial, with most of the additional suppression of subhalo abundance occurring with $t_{\rm first, infall} > 8$ Gyrs due to the increase in disk-to-halo mass ratio. These findings indicate that the MW disk predominantly loses early-infalling subhalos due to disk tidal stripping, and increasing the disk mass further enhances this effect.

\section{Discussion}
\label{sec:discussion}

\subsection{A user's guide to interpreting the scope of theoretical subhalo disruption/mass loss studies}
\label{discussion:interpretting}

Following on the discussion in Section \ref{sec:discussion:disruption}, we outline several principles for interpreting the results of studies like our own, which analyze simulated subhalos and depend on how many subhalos have lost so much mass that they have dropped out of the sample (a process often colloquially referred to as ``disruption''). We then outline where our current study fits relative to these points.

{\bf Quantitative vs. Qualitative}: The most important question in any subhalo study is whether its conclusions require that its results are {\em quantitatively} correct, or whether it is sufficient that the results are {\em qualitatively correct}. For example, a study claiming that observed satellite population statistics are in conflict with $\Lambda$CDM will almost always require quantitatively correct results, while for a study arguing that two quantities are correlated, qualitatively correct results may be sufficient. This distinction is important because quantitative studies require quantitative attention to the modeling points we bring up below, something that is quite difficult to do. In some cases, we consider such modeling to be an unsolved problem. Qualitative studies are not exempt from these concerns but can usually weather cruder modeling choices. 

{\bf Sample Selection}: When studying subhalo populations, one generally needs to select objects based on a mass proxy.\footnote{One could analyze all the objects output by a halo finder, but this is effectively still a mass selection, just one that is driven by numerical effects and may leave the analysis dominated by poorly resolved objects.} Popular choices are to select by present-day halo properties such as virial mass $M_{\rm vir}$ and maximum circular velocity $v_{\rm max}$, or to select by pre-infall mass proxies like $M_{\rm peak}$ or $v_{\rm peak}$, which are the historical maximum of $M_{\rm vir}$ and $v_{\rm max}$. 

Each selection choice has its own considerations. First, they target different classes of observations: present-day measures of mass are good tracers of observations that directly probe present-day subhalo mass (e.g., stream gaps, strong lensing, rotation curves, etc.; \citealt{2022arXiv220307354B}), while pre-infall proxies tend to be better tracers of satellite galaxies selected by stellar mass~\citep[e.g., ][]{2013ApJ...771...30R}. Second, population demographics can be very different for samples selected by pre-infall mass proxies than present-day mass proxies. This is particularly true for the radial distribution of subhalos, which is more concentrated in subhalos selected by peak mass than by present mass~\citep[e.g., ][]{2005ApJ...618..557N,2016MNRAS.457.1208H}. Third, numerical concerns are different for different mass proxies~\citep{2024ApJ...970..178M}: i) pre-infall selections have more stringent resolution requirements than present-day mass selections and may need to contend with aphysical fluctuations in mass as subhalos disrupt (see also \citealt{2017MNRAS.468..885V}); ii) present-day mass measurements need to contend with the fact that different subhalo finders have different ways of separating the mass of a subhalo from its host and different schemes can bias masses by tens-of-percent; iii) maintaining unbiased $v_{\rm max}$ values in subhalo velocity profiles for subhalos undergoing mass loss requires almost an order of magnitude more resolution than maintaining unbiased $M_{\rm vir}$ values.

{\bf Resolution}: A poorly resolved subhalo may not give accurate predictions for the $\Lambda$CDM model. As subhalos lose mass, they experience more classical two-body scattering~\citep[e.g., ][]{2003MNRAS.338...14P,2019MNRAS.488L.123L}; they may further experience decreased durability due to their simulation's force softening~\citep{2018MNRAS.475.4066V,2021MNRAS.500.3309M}, and can eventually encounter runaway mass loss due to discreteness effects~\citep{2018MNRAS.475.4066V}. In a well-calibrated simulation, all three effects affect subhalos at similar resolution levels~\citep{2024ApJ...970..178M}, but this may not be true in a simulation with excessively large or small force softening scales. Certain subhalo population statistics are strongly dependent on resolution, such as the abundance or radial distribution of subhalos selected at a fixed $M_{\rm peak}$.

{\bf Subhalo finder}: Subhalo finders are imperfect tools and are often the dominant source of numerical biases in subhalo analysis. For example, the \textsc{Rockstar} subhalo finder~\citep{2013ApJ...762..109B} is one of the most widely used subhalo finders and generally performs at least as well as other major subhalo finders in comparative tests between tools~\citep{2011MNRAS.415.2293K,2012MNRAS.423.1200O,2013MNRAS.429.2739O,2013MNRAS.436..150S,2014MNRAS.441.3488A,2014ApJ...787..156B,2019PASA...36...28E}. However, \textsc{Rockstar} demonstrably loses track of even high-resolution subhalos after relatively modest amounts of mass loss~\citep{2016ApJ...818...10G,2024ApJ...970..178M} and falsely converges~\citep{2024ApJ...970..178M}, meaning that even as resolution is increased and the simulation's subhalos are able to survive for longer, \textsc{Rockstar} will continue losing track of subhalos after they have lost the same amount of mass. The strengths and drawbacks of every halo finder are different, and it is important to understand what those weaknesses are in depth, even for qualitative studies. For some subhalo finders, these limitations have never been formally studied. We also recommend that one should generally be skeptical of reliability arguments based on subhalo finder comparisons or internal convergence tests, given that such tests have been performed extensively on the \textsc{Rockstar} subhalo finder without identifying its aforementioned issues.

{\bf Galaxy Mass Loss/Orphan Model} To first order, particles can be stripped from a subhalo when they orbit close to that subhalo's tidal radius, and a subhalo's tidal radius gradually shrinks as it orbits its host. Because galaxies are small compared to their halos, low-mass satellite galaxies generally stay intact through their initial mass loss and begin to rapidly lose mass after passing a critical threshold in $M_{\rm vir}/M_{\rm peak}$~\citep{2013MNRAS.429.1066S}. Whether or not this rapid mass loss is qualitatively consistent with ``disruption'' depends on the stellar density profile of the satellite~\citep{2023MNRAS.519..384E}. For subhalo finders that lose track of subhalos after this threshold is reached \citep[e.g. \textsc{Symfind}, ][]{2024ApJ...970..178M}, a galaxy mass loss model is needed, lest the finder overpredict the abundance of satellite galaxies at a fixed stellar mass. For subhalo finders that lose track of subhalos prior to this threshold \citep[e.g. \textsc{Rockstar}][]{2024ApJ...970..178M} require both a galaxy disruption model and an ``orphan'' model, i.e., a model that creates post hoc galaxy tracers and evolves them forward based on heuristics. Several points to consider when evaluating these models are (a) that mass loss models calibrated on hydrodynamic simulations likely vastly overestimate mass loss rates due to numerical issues~\citep[e.g., ][]{2019MNRAS.488L.123L}, (b) that dynamical friction is quite weak in low-mass subhalos~\citep{2016MNRAS.455..158V} meaning that it is unlikely to be their primary disruption mechanism despite it being a popular prescription, and (c) that the popular choice of generating orphan galaxies during a subhalo's last snapshot is questionable given that the subhalo recovered during this snapshot is often extremely unreliable.

With this context, let us consider the current study. We are studying $M_{\rm peak}$-selected subhalo samples. The goal of this work is to use these $M_{\rm peak}$ selected samples to qualitatively emulate the behavior of stellar-mass selected samples. We do this because we consider the work associated with producing a quantitatively reliable model of stellar mass loss to be beyond the scope of this study. We consider this a reasonable approximation at this level of accuracy because the majority of our subhalos have $M_{\rm vir}/M_{\rm peak}$ high enough ($\gtrsim 0.2$) that they are unlikely to have started galaxy disruption (see Section \ref{sec:discussion:disruption}).

The interplay between the numerics of our simulations and our chosen subhalo finder, \textsc{Symfind}, has been extensively explored in \citet{2024ApJ...970..178M}. \textsc{Symfind} is able to track subhalos well past the point where their galaxies have started to rapidly lose mass and will not be a limiting factor in analysis. However, some subhalos in the $n_{\rm peak} \lesssim 5\time 10^3$ range will experience some degree of accelerated mass loss due to simulation resolution prior to the point when stellar mass loss should begin, although we do not currently have a model that can make quantitative estimates for how many subhalos will be missing in this range. The use of \textsc{Symfind} significantly increases the number of subhalos that we find relative to \textsc{Rockstar} (see Appendix \ref{sec:app:res}), and the choice between subhalo finders is comparable in effect to the amount of suppression introduced by the disk itself. Much of this difference is due to \textsc{Rockstar} losing track of subhalos early, but some unknown fraction of the difference may come from the fact that \textsc{Symfind} is holding onto subhalos so long that their satellite galaxies would have lost enough stellar mass to move outside the equivalent stellar-mass-selected sample. Future work on precise stellar mass loss models will help resolve this issue.

\subsection{Comparisons with previous embedded disk simulations}
\label{sec:discussion:discrepancies}

To begin with, conclusions on subhalo abundance suppression as described in \citet{2017MNRAS.471.1709G} and \citet{2019MNRAS.487.4409K} are largely {\em consistent} with the two EDEN halos with $M_{\ast, \rm Disk}/M_{\rm vir, host}>0.05$ and the EDEN $M_{\ast, \rm Disk}\times 2.5$ suite that have MW-like heavy disk-to-halo mass ratios (Section~\ref{sec:results}). Both simulation works have $\sim 10\times$ higher numerical resolution than EDEN and presented disk effects for subhalos that reached an order-of-magnitude lower masses than our $M_{\rm peak} \geqslant 1.2\times 10^8\mathrm{M_{\astrosun}}$. Therefore, we qualitatively compare these previous simulations with EDEN and emphasize the subhalo count ratios from \citet{2017MNRAS.471.1709G}, SPMF figure.  

In Phat-ELVIS~\citep{2019MNRAS.487.4409K}, $\sim 35\%$ subhalos within 100 kpc are lost down to the EDEN resolution limit (see their Fig. 2, peak maximum circular velocity $v_{\rm peak}\geqslant10\mathrm{\,km\,s^{-1}}$), which is consistent with EDEN (Fig.~\ref{fig:Nratio}). However, subhalo abundance suppression for one of the two halos in \citet{2017MNRAS.471.1709G} is significantly stronger than EDEN (Fig.~\ref{fig:Nratio} top panel, data taken from their Fig. 4 at $M_{\rm peak} = 1.2\times 10^8 \mathrm{M_{\astrosun}}$). This host in \citet{2017MNRAS.471.1709G} has a {\em lower} $M_{\ast, \rm Disk}/M_{\rm vir, host}$ than its sibling but is left only with $30\%$ of subhalos within 100 kpc relative to its DMO run. This is counterintuitive, as its disk formed later than its sibling and always had smaller $M_{\ast}$. It is very rare for a MW-mass halo in EDEN to lose $>60\%$ of its subhalos, even when considering the largest $M_{\ast, \rm Disk}/M_{\rm vir, host}$ hosts in EDEN $M_{\ast, \rm Disk}\times 2.5$ with $\sim 2\times$ larger disk-to-halo mass ratios. 

We conjecture two possible factors that can potentially lead to the stronger subhalo abundance suppression seen in \citet{2017MNRAS.471.1709G}. One possibility is that the subhalo orbits of these two halos had smaller pericenters and were accreted more radially due to the large-scale environment of the host (e.g., accreting subhalos collectively from filaments). Another possibility is that the fragility of AHF halo finder used by \citet{2017MNRAS.471.1709G} has even worse subhalo tracking issues than \textsc{Rockstar}~\citep{2024ApJ...970..178M}, leading to more heavily stripped subhalos being artificially lost track of in the presence of a disk potential. A similar effect could be applicable to \textsc{AHF}~\citep{2009ApJS..182..608K}, the halo finder used in \citet{2017MNRAS.471.1709G}, such that it numerically enhances the suppression of subhalo abundance. Since only two halos were picked from FIRE-2 and they already happen to yield drastically lower $N_{\rm Disk}/N_{\rm DMO}$, the average mass loss for subhalos in FIRE-2 is likely to be statistically stronger than that of EDEN (combining physical and numerical effects). We defer a more careful analysis of this issue to future work.

Regarding subhalo radial distributions, EDEN qualitatively agrees with \citet{2017MNRAS.471.1709G} and \citet{2019MNRAS.487.4409K}. However, there are a few quantitative differences we wish to highlight. Comparing EDEN with \citet{2017MNRAS.471.1709G}, the latter (see their Fig. 3) has a similar 3D radial distribution to EDEN $M_{\ast, \rm Disk}\times 2.5$ (Fig.~\ref{fig:radial} left panel) and loses $>2/3$ subhalos within 30 kpc due to the added disk. The differences in 3D subhalo distances mainly come from the cumulative subhalo counts within 100 kpc being $\sim 15\%$ lower than EDEN $M_{\ast, \rm Disk}\times 2.5$ following the discussion above. Comparing Phat ELVIS to EDEN, both Phat ELVIS and EDEN $M_{\ast, \rm Disk}\times 2.5$ do not have subhalos within 20 kpc of their hosts, with the suppression ratio decreasing at increasing distances (see their Fig. 3). 

The more significant differences between EDEN and these works are the pericenter distance distributions. Symphony DMO has $\lesssim 15\%$ subhalos with $d_{\rm peri}<10$ kpc on average, and in EDEN $30\%\sim 40\%$ of subhalos within 10 kpc are lost due to the added disk. However, the two halos in \citet{2017MNRAS.471.1709G} have $\gtrsim 20\%$ subhalos within 10 kpc and reach $50\%$ at $d_{\rm peri}\sim 20$ kpc in their DMO run, having much tighter pericenters than EDEN to start with. In their Disk runs (see their Fig. 5), there are no subhalos with $d_{\rm peri} <10$ kpc, but $10\%$ of EDEN subhalos have $d_{\rm peri}\lesssim 10$ kpc. They show that the subhalo density increases monotonically with smaller $d_{\rm peri}$ in the DMO case and significantly drops in the Disk run within $d_{\rm peri}\sim 35$ kpc. This point of subhalo count suppression mostly occurring at $d_{\rm peri}\lesssim 35$ kpc is also reproduced by the Phat ELVIS simulations, which have similar numerical resolution but are based on a different halo finder (their Fig. 4). The orbital interpolation cadence of 10 Myrs adopted in EDEN is similar to that in \citet{2017MNRAS.471.1709G} and \citet{2019MNRAS.487.4409K}, also unlikely to be a major factor creating the differences in pericenter distributions. 

Given that these two previous simulations reach similar pericenter distributions with different halo finders, it is most likely that the higher numerical resolution adopted in them leads to both having way more subhalos in their DMO runs with $d_{\rm peri}<20$ kpc compared to SymphonyMilkyWay (DMO). The higher number density in their DMO runs sets a higher $N_{\rm DMO}$, leading to a much lower $N_{\rm Disk}/N_{\rm DMO}$ compared to EDEN EDEN. As mentioned in Section~\ref{sec:results:radial}, the subhalo radial distributions are sensitive to numerical resolution, and perhaps the pericenter distances are particularly more sensitive since they probe the most tidally violent part of each subhalo's orbit that requires the most resolving power to numerically converge~\citep{2010MNRAS.406.1290P,2018MNRAS.475.4066V,2018MNRAS.474.3043V,2020MNRAS.491.4591E,2021MNRAS.505...18E,2024ApJ...970..178M}. Please refer to Appendix~\ref{sec:app:res} for a discussion of resolution effects. 

Finally, EDEN results are also largely consistent with \citet{2022MNRAS.509.2624G} in which various properties of the disk impact on NFW-profile (cuspy) subhalos are explored using the SatGen semi-analytic model~\citep{2021MNRAS.502..621J}. They found that the disk mass is the most dominant factor in determining subhalo abundance suppression intensity over other factors such as disk scale radius and axis ratio. Within their framework, they find that heavier disks cause stronger subhalo abundance suppression, with $M_{\ast, \rm Disk}/M_{\rm vir, host}=0.05$ disks causing $\sim 30\%$ subhalos lost within $50$ kpc and $\sim 10\%$ subhalos lost within $R_{\rm vir, host}$, qualitatively agreeing with the EDEN results (Fig.~\ref{fig:Nratio}). They also find that the radial distribution of subhalos is more suppressed in the inner regions within $\sim 100$ kpc with increasing disk mass, similar to EDEN (Fig.~\ref{fig:radial}). These findings further corroborate the key role disk-to-halo mass ratio plays in modeling the tidal stripping and mass loss of subhalos due to the central galaxy. 

We note that in the case of cored-subhalos due to internal stellar feedback as in hydrodynamic simulations~\citep[e.g., ][]{2017MNRAS.471.1709G,2020MNRAS.491.1471S} or emulated semi-analytic models~\citep[e.g., ][]{2021MNRAS.502..621J}, subhalo abundance suppression could be more intense due to the less-concentrated nature of the subhalo itself as compared to the cuspy DMO subhalos. This effect is best exemplified in the two halos from \citet{2017MNRAS.471.1709G} where the subhalo abundance within 100 kpc in their hydrodynamic runs is always lower than their embedded disk runs. Therefore, more subhalo mass loss and abundance suppression could occur in reality than the EDEN predictions in this work.

\subsection{Caveats with fixing the disk orientation}
\label{sec:discussion:disk}

As mentioned in Section~\ref{sec:method:disk}, we fix the disk potential axis to the MW host halo spin at $z=0$ throughout the simulation. This choice follows previous arguments that the specific shape of the disk potential does not change the quantitative results of subhalo abundance suppression~\citep{2017MNRAS.471.1709G,2022MNRAS.509.2624G} and has been implemented similarly in Phat ELVIS~\citep{2019MNRAS.487.4409K}. However, this simplistic treatment means that the torque exerted by the disk potential on the subhalos is not back-reacted on the disk, which could otherwise cause the disk to precess and warp. Without this precession, embedded disk simulations set a preferred direction of collapse at halo centers and can create a more ellipsoidal central dark matter density distribution ($\lesssim 10$ kpc) than their hydrodynamic counterparts (e.g., Fig. 1 lower panel in \citealt{2017MNRAS.471.1709G}).

The torque on the subhalos due to the assumed rigidity in disk direction may cumulatively influence the spin of the disk that may not keep it aligned with the host halo spin. Additional external torques from nearby structures outside the MW host may also contribute in addition to the subhalos, eventually concertedly shifting the 3D trajectory of the disk. Indeed, as we show in Appendix~\ref{sec:app:traj}, the EDEN hosts end up at locations that are on average $\gtrsim15$ kpc from their Symphony DMO counterparts at $z=0$, sometimes even deviating at $\sim 40$ kpc. These shifts in the host halo center do not have significant impacts on our conclusions as the disk (sink particle) is kept within 0.1 kpc of the halo center (Fig.~\ref{fig:traj}) and the subhalo infall times are also largely unchanged as visible through the jumps in host mass growth histories (Fig.~\ref{fig:MvirH}). This is consistent with the isolation criteria of SymphonyMilkyWay hosts as they are selected to be the most massive object within $4R_{\rm vir, host}$~\citep{2015ApJ...810...21M,2023ApJ...945..159N}, which makes the $\lesssim 0.1 R_{\rm vir, host}$ shifts in their locations unlikely to cause significantly different infalling subhalo populations. 

These effects indicate that although the disk direction fixture causes an artificial shift in halo trajectories and more ellipsoidal central densities, its impact on subhalo abundance suppression is negligible. The rigid disk assumption also represents an upper bound on the pure gravitational impact of the disk on subhalos, as the disk in our case cannot absorb orbital energy through precession or warping. This finding is also in line with numerical experiments in \cite{2024MNRAS.527.8841S}, which indicate that the orientation of the disk, the density profile of the disk (axisymmetric or spherical), and sizes all have little impact on subhalo orbits and the strength of subhalo abundance suppression.

\subsection{Effects of numerical resolution}
\label{sec:discussion:resolution}

As mentioned in Section~\ref{sec:results:radial}, the quantitative results in this work, such as the subhalo count ratio, SPMF, and radial distributions, may be subject to change under different numerical resolutions. We show in Appendix~\ref{sec:app:res} that, indeed, resolution matters, as well as the specific choice of the subhalo finder. We re-simulate five SymphonyMilkyWay hosts at $8\times$ higher resolution. Higher resolution leads to a higher SPMF that boosts subhalo counts at all mass scales. The boost in subhalo counts due to higher numerical resolution in the dark-matter-only case is comparable to the magnitude of disk stripping effects on subhalo abundances at the same fiducial resolution, further highlighting that the results of disk effects are sensitive to numerical resolution. 

The particle-tracking-based \textsc{Symfind} halo finder performs much better at tracking heavily stripped subhalos than the phase space overdensity-based halo finder \textsc{Rockstar} (see Section 4.2 and Fig. 5 in \citealt{2024ApJ...970..178M}). The key takeaway is that \textsc{Rockstar} misses a sizable fraction of $\lesssim 10^5$ peak-particle-count subhalos \citep[Fig. 10, ][]{2024ApJ...970..178M}, and the majority of the subhalos that Rockstar misses are heavily stripped. As a result, \textsc{Rockstar} yields false numerical convergence for the abundance of subhalos with $\sim 100$ to $10^5$ peak particles. The cause of this is rooted in the nature of the phase-space overdensity becoming less prominent as a subhalo starts being stripped; these overdensities thus become much harder to identify in \textsc{Rockstar} than the most-bound particles used in \textsc{Symfind}. This problem is likely even more severe for physical-overdensity-based halo finders such as \textsc{Subfind}~\citep{2001MNRAS.328..726S,2009MNRAS.399..497D}, where only spatial density is used to identify subhalos (see also the discussion in Appendix H of \citealt{2024ApJ...970..178M}). 

For heavily stripped subhalos in our fiducial-resolution EDEN simulations, we show that \textsc{Symfind} recovers more $\gtrsim 10^4$ peak particle count subhalos than \textsc{Rockstar} when resolution is increased (Appendix \ref{sec:app:res}); this result is consistent with Fig. 10 of \citet{2024ApJ...970..178M}. This is a regime where previous studies have long believed that phase-space and spatial overdensity halo finders are complete and converged~\citep[e.g.,][]{2003MNRAS.338...14P,2004MNRAS.353..624D,2005MNRAS.357...82R,2008MNRAS.391.1685S,2010MNRAS.402...21N,2012MNRAS.425.2169G,2019MNRAS.488.3663L,2021MNRAS.500.3309M,2023ApJ...945..159N}, whereas they are in fact underestimating subhalo abundances compared to idealized simulations~\citep{2010MNRAS.406.1290P,2018MNRAS.474.3043V,2018MNRAS.475.4066V, 2020MNRAS.491.4591E,2021MNRAS.505...18E}. 

Nevertheless, we emphasize that the EDEN results---and particularly the $N_{\rm Disk}/N_{\rm DMO}$ predictions---are more robust than previous studies given our current numerical resolution. This follows as we focus on \emph{relative} subhalo abundances in disk versus DMO runs rather than absolute abundances, paired with particle-tracking subhalo finder \textsc{Symfind}. As we have checked, the subhalo count ratios shown in Fig.~\ref{fig:Nratio} are more consistent between \textsc{Symfind} and \textsc{Rockstar} than the SPMF (Fig.~\ref{fig:SHMF_conv}) which is more sensitive to the specific halo finder choice. We suggest that future work which includes quantitative comparisons to EDEN results should control for numerical resolution and subhalo finder differences. Qualitative comparisons between different simulations using different halo finders should prioritize subhalo abundance \emph{ratio} ($N_{\rm Disk}/N_{\rm DMO}$) comparisons rather than absolute subhalo abundance (SPMF).

\subsection{The Milky Way's disk in a cosmological context}
\label{sec:discussion:MW}

The most important finding in this work is that subhalo abundance suppression in MW-mass halos is sensitive to the disk-to-halo mass ratio, and thus the MW (as well as M31), with a larger-than-average disk, likely suppresses subhalos more efficiently (Fig.~\ref{fig:Nratio}). This has profound implications for understanding the galaxy--halo connection of low-mass satellites around MW-mass hosts external to the MW. If one tries to constrain the galaxy--halo connection on a broad set of MW-mass objects in the Local Universe from Surveys such as SAGA~\citep{2017ApJ...847....4G,2021ApJ...907...85M,2024ApJ...976..117M,2024ApJ...976..118G,2024ApJ...976..119W} and ELVES~\citep{2022ApJ...927...44C}, the overall effect of subhalo abundance suppression across their host samples would be weaker than in the MW itself, due to the rareness ($\sim 1\sigma$ outlier) of a MW-like heavy stellar disk. 

For instance, the MW is a $\gtrsim1\sigma$ larger-than-average host compared to the 101 hosts in SAGA (see Fig. 2 in \citealt{2024ApJ...976..119W}), consistent with EDEN hosts shown in Figs.~\ref{fig:smhm} and \ref{fig:Nratio}. For the majority of SAGA hosts, which coincide with $M_{\ast, \rm Disk}/M_{\rm vir, host}\lesssim 0.02$, disk effects of $\sim 10\%$ within $R_{\rm vir, host}$ and $\sim 20\%$ within 100 kpc are expected for subhalos according to EDEN (Fig.~\ref{fig:Nratio}). In those cases, constraints on the low-mass galaxy--halo connection based on average number densities of satellites within $R_{\rm vir, host}$ are rather robust given the expected $\sim 10\%$ disk effects, but radial trends could change significantly especially within the central 100 kpc due to stronger suppression at smaller radii ($\gtrsim 20$, Fig.~\ref{fig:radial}). This indicates that EDEN might mitigate the limitation of UM-SAGA, which predicts fewer quenched satellites than SAGA observations within 100 kpc \citep{2024ApJ...976..119W}, by accounting for the additional tidal forces from the host galactic disk.

However, if one focuses exclusively on low-mass satellites in the MW itself (e.g. \citealt{2020ApJ...893...48N}), modeling subhalo abundance suppression properly becomes crucial due to the presence of a massive stellar disk that is more effective at stripping subhalos. In fact, \citet{2020ApJ...893...48N} showed that the subhalo abundance suppression intensity \footnote{Implemented using the random forest model introduced in \citet{2018ApJ...859..129N}.} of \citet{2017MNRAS.471.1709G} is consistent with the luminosity, size, and spatial distribution of MW satellite galaxies but is not precisely constrained by the current data (though future surveys will help reduce uncertainties, e.g. see \citealt{2024ApJ...967...61N}). Therefore, our embedded disk simulations, especially the EDEN $M_{\ast, \rm Disk} \times 2.5$ subsample, establish a novel platform for more realistically modeling subhalo abundances that is particularly important for interpreting the MW. 

In this context, it is interesting to compare the $M_{\ast, \rm Disk}/M_{\rm vir, host}$ ratios and satellite counts in the MW and M31, which have 5 and 8 satellites with $M_{\ast}> 10^7 \mathrm{M_{\astrosun}}$, respectively~\citep{2012AJ....144....4M,2020ApJ...893...47D}. According to Fig.~\ref{fig:Nratio}, M31 has a higher $M_{\ast, \rm Disk}/M_{\rm vir, host}$ than the MW and should thus have a lower $N_{\rm disk}/N_{\rm DMO}$ due to stronger tidal stripping. Some results suggest that the MW~\citep[$1.08^{+0.14}_{-0.20}\times 10^{12}\mathrm{M_{\astrosun}}$][]{2020MNRAS.494.4291C} and M31~\citep[$1.4^{+0.4}_{-0.4}\times 10^{12}\mathrm{M_{\astrosun}}$][]{2010MNRAS.406..264W} have similar halo masses, although the uncertainties allow for MW/M31$\sim 1:2$ virial mass ratios.\footnote{The Local Group ``timing argument'' also allows for a significantly larger M31 virial mass, e.g., \citealt{2024A&A...689L...1B}.} Nevertheless, given the similar mean-inferred halo masses, we should also consider secondary halo bias and intrinsic host-to-host scatter in subhalo counts at fixed halo mass, which is embedded in the scatter of $N_{\rm Disk}/N_{\rm DMO}$ in Fig.~\ref{fig:Nratio}. As shown in the SAGA Survey Paper III~\citep{2024ApJ...976..117M}, the observed satellite count in MW-mass hosts is predominantly correlated with the stellar mass of their most massive satellite (Fig. 15). Recent observations~\citep{2021MNRAS.504.5270D} indicate that M31's most massive satellite M32 could have a stellar mass as large as $2.5\times 10^{10}\mathrm{M_{\astrosun}}$, making it almost an order-of-magnitude heavier than the LMC~\citep[$2.7 \times 10^9 \mathrm{M_{\astrosun}}$,][]{2002AJ....124.2639V}. This could mean that M31 had a much higher $N_{\rm DMO}$ to start with, even if its virial mass is comparable to the MW. In this scenario, when M31's subhalos go through much stronger tidal stripping due to its heavier disk, M31 still ends up having a higher $N_{\rm disk}$ than the MW while maintaining a lower $N_{\rm Disk}/N_{\rm DMO}$. 

Generalizing the disk impact from MW-mass hosts to other host mass scales, we expect subhalo abundance suppression due to disk tidal stripping to be weaker in other halo mass scales given that the stellar-to-halo mass ratio is higher around $M_{\rm vir, host}\sim 10^{12}\mathrm{M_{\astrosun}}$ MW-mass halos~\citep[e.g., ][]{2013ApJ...770...57B,2019MNRAS.488.3143B,2023ApJ...945..159N,2024ApJ...976..119W}. However, subhalo abundance in group-mass halos $M_{\rm vir, host}\sim 10^{13}\mathrm{M{\astrosun}}$ might be affected by MW-mass progenitors whose pre-processed~\citep{2015ApJ...807...49W} subhalo abundance might carry over. This is an interesting topic to be explored in the future, but we speculate that it might be highly dependent on the specific assembly history of the group host, i.e., at what redshift those MW-sized satellites fell in, and whether their galaxies grew massive enough to significantly affect their satellite population at their accretion redshift. Dedicated zoom-in simulations embedding galaxies into both the \textsc{Symphony} group-mass hosts and their MW-mass infall progenitors would be required to answer these questions.

In the near future, we plan to extend the EDEN framework to the new `MW-est' zoom-in simulations~\citep{2024ApJ...971...79B}, a suite of 20 MW-like zoom-in simulations that have similar halo masses as SymphonyMilkyWay hosts but also have tailored Gaia-Sausage Enceladus~\citep{2018MNRAS.478..611B,2018Natur.563...85H} and LMC~\citep{2002AJ....124.2639V,2007ApJ...668..949B,2013ApJ...764..161K,2023Galax..11...59V} mergers.

\section{Conclusions}
\label{sec:conclusion}

In this paper, we introduced the \textsc{EDEN}-Symphony simulation suite, a set of 45 MW-mass dark-matter-only (DMO) zoom-in simulations with analytic disk potentials embedded in the host halo centers to capture subhalo abundance suppression due to tidal stripping and mass loss under the influence of a central galaxy baryonic disk. The baseline DMO zoom-in simulations of these 45 MW-mass halos are part of the Symphony compilation~\citep{2023ApJ...945..159N}. \textsc{EDEN}-Symphony enlarges the number of existing zoom-in simulations with embedded disk potentials in the literature by a factor of three. 

A novel feature of the work is that the analytic disk potentials grow self-consistently with their equivalent stellar masses predicted by the empirical galaxy--halo connection model \textsc{UniverseMachine}~\citep{2019MNRAS.488.3143B} (Section~\ref{sec:method}). This ensures that the broad range of disk masses and disk formation histories are well-sampled in a correlated fashion with halo mass and halo assembly. Our simulation suite focuses on $M_{\rm vir}\sim 10^{12.1}\mathrm{M_{\astrosun}}$ halos and covers heavy disk systems like our MW (Fig.~\ref{fig:sfh}). We use the particle-tracking subhalo finder \textsc{Symfind} \citep{2024ApJ...970..178M} to track subhalos down to $M_{\rm peak}\geqslant 300m_{\rm DM}=1.2\times 10^8\mathrm{M_{\astrosun}}$ and quantify how the subhalo populations change in the presence of a central disk potential. We have also re-simulated nine embedded-disk halos with $\times 2.5$ heavier disks (EDEN $M_{\ast, \rm Disk}\times 2.5$) to explore the effect of disk mass while fixing the disk growth history. Our main findings are:
\begin{enumerate}
    \item Subhalo abundance suppression due to the disk is sensitive to the host's $M_{\ast, \rm Disk}/M_{\rm vir, host}$ such that more suppression occurs in hosts with larger $M_{\ast, \rm Disk}/M_{\rm vir, host}$ (Figs.~\ref{fig:map} and \ref{fig:Nratio}).

    \item Among MW-mass halos ($M_{\rm vir, host}\sim 10^{12}\mathrm{M_{\astrosun}}$), the MW and M31 ($M_{\ast, \rm Disk}/M_{\rm vir, host}\gtrsim 0.05$) have particularly strong subhalo abundance suppression ($\gtrsim 30\%$ within 100 kpc, Fig.~\ref{fig:Nratio}) due to their $\gtrsim 2\sigma$ up-scatter in stellar mass (Fig.~\ref{fig:smhm}) and early disk formation (Fig.~\ref{fig:sfh})

    \item In EDEN fiducial hosts, subhalos with large peak masses ($M_{\rm peak}\gtrsim 10^{10}\mathrm{M_{\astrosun}}$) are more likely to be lost than lower-mass counterparts with the inclusion of the disk. In the more massive disks of EDEN $M_{\ast, \rm Disk}\times 2.5$, however, subhalo abundance suppression is independent of subhalo peak mass (Fig.~\ref{fig:SHMF}). 

    \item Subhalos with small pericenters ($d_{\rm peri} < 100$ kpc, Fig.~\ref{fig:radial}) and early infall times (lookback time $\gtrsim 8$ Gyrs, Fig.~\ref{fig:tinfall}) are preferentially lost.

    \item We advocate that comparisons between different simulations should prioritize comparing subhalo abundance \emph{ratio} ($N_{\rm Disk}/N_{\rm DMO}$) rather than subhalo mass functions to mitigate subhalo finder systematics.
\end{enumerate}

The most important findings of this work are the first two points mentioned above. It is clear from Figs.~\ref{fig:smhm} and \ref{fig:Nratio} that the MW has an atypically large disk for an average $M_{\rm vir}\sim 10^{12}\mathrm{M_{\astrosun}}$ halo. Previous simulation work~\citep{2010ApJ...709.1138D,2017MNRAS.471.1709G,2019MNRAS.487.4409K} with embedded disk potentials implementing MW-like heavy stellar disks ($M_{\ast}\gtrsim 6\times 10^{10}\mathrm{M_{\astrosun}}$) is potentially biased towards heavy disk systems and overestimates the average subhalo abundance suppression intensities in MW-mass halos. Subhalo abundance suppression is less significant in a broader range of MW-mass halos in the Universe, and suppression ratios obtained in previous models using heavy disks should not be naively generalized to the broader range of MW-mass hosts at $M_{\rm vir}\sim 10^{12}\mathrm{M_{\astrosun}}$. The disk effects on subhalo abundances at {\em other host halo mass scales} significantly smaller or larger than $M_{\rm vir, host}\sim 10^{12}\mathrm{M_{\astrosun}}$ are likely even less important due to their smaller $M_{\ast, \rm Disk}/M_{\rm vir, host}$~\citep{2018ARA&A..56..435W}.

\section*{Acknowledgments}

We thank James Bullock, Benedikt Diemer, Pratik Gandhi, Yao-Yuan Mao, Volker Springel, Frank van den Bosch, Andrew Wetzel and the GFC group at KIPAC/Stanford for helpful discussions and insightful comments during the preparation of this draft. The FIRE-2 cosmological zoom-in simulations of galaxy formation are part of the Feedback In Realistic Environments (FIRE) project, generated using the Gizmo code~\citep{2015MNRAS.450...53H} and the FIRE-2 physics model~\citep{2018MNRAS.480..800H}.

This work was supported in part by NASA through HST-AR-17044, by the National Science Foundation under Grant No. NSF PHY-1748958, by the U.S. Department of Energy under contract number DE-AC02-76SF00515 to SLAC National Accelerator Laboratory. 
DY and HBY were supported by the John Templeton Foundation under grant ID \#61884 and the U.S. Department of Energy under grant number DE-SC0008541.
Part of this research was conducted during the GALEVO-23 KITP workshop, supported in part by grant NSF PHY-1748958 to the Kavli Institute for Theoretical Physics (KITP). We thank the organizers of the KITP workshop; discussions during this meeting were helpful in developing this research.
This research made use of computational resources at SLAC National Accelerator Laboratory, a U.S.\ Department of Energy Office of Science laboratory and the Sherlock cluster at the Stanford Research Computing Center (SRCC); the authors are thankful for the support of the SLAC  and SRCC computational teams. This work used the Extreme Science and Engineering Discovery Environment (XSEDE) Stampede2 and Ranch systems at the Texas Advanced Computing Center (TACC) through allocation TG-PHY210112. The authors acknowledge TACC at The University of Texas at Austin for providing HPC resources that have contributed to the research results reported within this paper.

\vspace{5mm}
\facilities{Stanford Research Computing Center, SLAC Data Facility, TACC Stampede2}

\software{Astropy~\citep{2013A&A...558A..33A,2018AJ....156..123A,2022ApJ...935..167A},  Matplotlib~\citep{2007CSE.....9...90H}, NumPy~\citep{harris2020array}, SciPy~\citep{2020SciPy-NMeth}, arXiv, ADS}

\bibliography{disk}{}
\bibliographystyle{aasjournal}

\appendix
\label{sec:app}

\section{Disk dynamics and density distribution}
\label{sec:app:disk}

\begin{figure*}
    \centering
	\includegraphics[width=2\columnwidth]{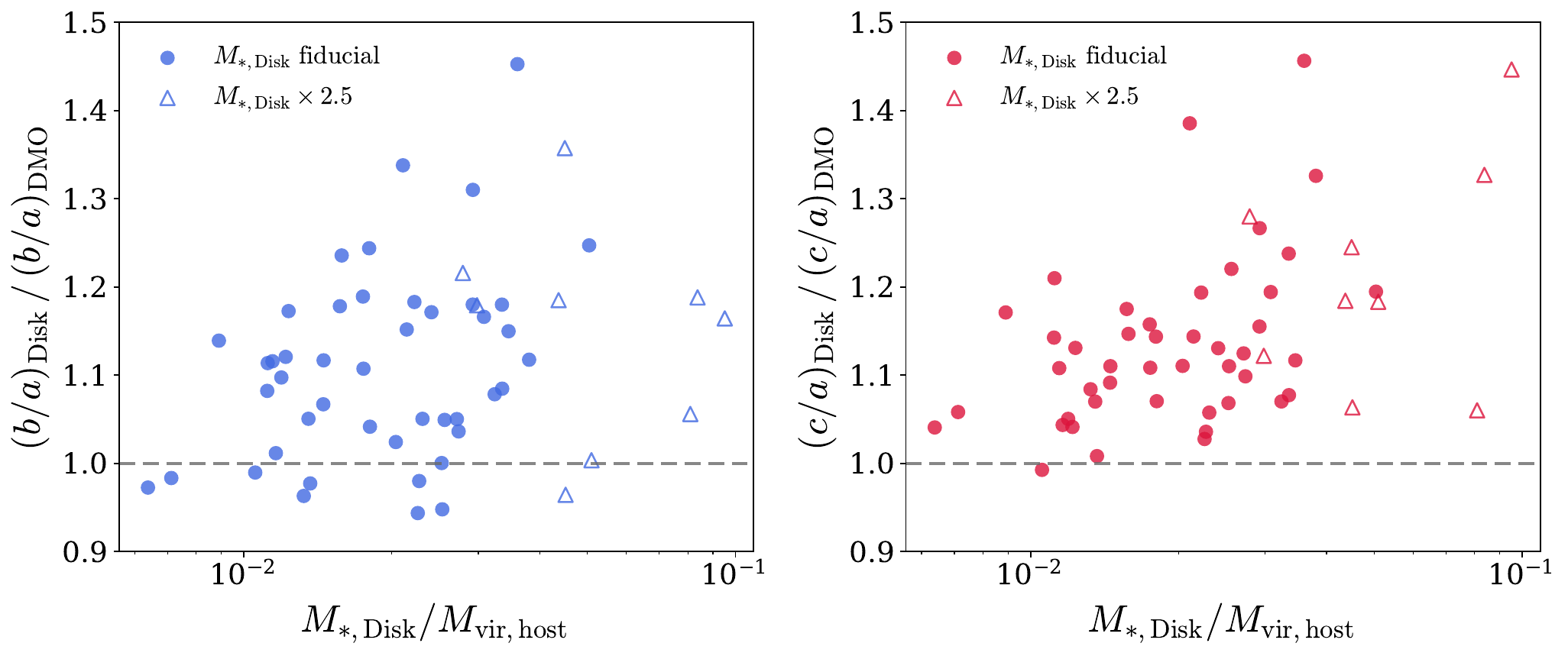}
    \caption{Semi-medium ($b$) and semi-minor ($c$) axis ratios with semi-major axis ($a$) of the MW host halo. The  axis ratios lengths are determined by the moment of inertia for dark matter particles within $R_{\rm vir, host}$. The left panel shows the $b/a$ ratios with and without the embedded disk potential, and right panel shows the $c/a$ ratios. Solid point denote the 45 host with fiducial UM stellar masses, while the open triangles are the nine hosts ran with $\times 2.5$ heavier disks. With larger $M_{\ast, \rm Disk}/M_{\rm vir, host}$, host halos tend to become rounder and have larger axis ratios. These two positive correlations have Spearman $r_S=0.30$ ($b/a$) and $r_S=0.46$ (c/a).}
    \label{fig:axis}
\end{figure*}

Following the disk potential implementation in Section~\ref{sec:method:disk}, we provide in this Appendix a more detailed derivation of the dynamics for the disk-sink particle combination, the disk acceleration on dark matter particles, and disk equivalent density given the disk potential.

The dynamics of the disk--sink are determined by force contributions from two parts. Part one is the N-body forces from all other dark matter particles on the sink particle:
\begin{equation}
\label{eq:A1}
    \vec{F}_{\rm sink} = \sum_{i=1}^{N_{\rm part}} m_{i} \vec{a}_{\mathrm{sink}, i} = \sum_{i=1}^{N_{\rm part}} m_{i} \frac{GM_{\rm sink}}{|r_i|^3} \vec{r}_{i}.
\end{equation}
Here, $N_{\rm part}$ is the total number of particles in the simulation, $M_{\rm sink}$ is the sink particle mass, $m_i$ is the particle mass of the $i$-th particle,  $\vec{r}_i$ is the position vector of the $i$-th particle relative to the sink particle, and $\vec{a}_{\mathrm{sink}, i}$ is the acceleration caused by the $i$-th particle on the sink particle.

Part two is the back reaction of forces exerted by the disk potential on all other particles. If we assume the disk is {\em spherically symmetric} with equivalent mass $M_{\rm disk}$, then the back reaction of all other particles on the disk is:
\begin{equation}
    \vec{F}_{\rm Sph} = \sum_{i=1}^{N_{\rm part}} m_{i} \vec{a}_{\mathrm{Sph}, i}\  = \ \sum_{i=1}^{N_{\rm part}} m_{i} \frac{GM_{\rm disk}}{|r_i|^3} \vec{r}_{i}.
\end{equation}
Here, $\vec{a}_{\mathrm{Sph}, i}$ is the acceleration on the spherical potential due to the gravitational interaction with the $i$-th particle. The subscript `Sph' in the vectors stands for `Spherical'.

However, since the disk potential $\Phi(R,Z)$ is axisymmetric and not spherically symmetric, the back reactions on the disk from most particles  {\em do not point towards the disk center}, which is the sink particle location. This means that the second equality does not hold and:
\begin{equation}
    \vec{F}_{\rm disk} = \sum_{i=1}^{N_{\rm part}} m_{i} \vec{a}_{\mathrm{disk}, i}\  \neq \ \sum_{i=1}^{N_{\rm part}} m_{i} \frac{GM_{\rm disk}}{|r_i|^3} \vec{r}_{i},
\end{equation}
and the disk \emph{cannot} be treated as a point mass. 

Thus, the total acceleration on the disk potential-sink particle combination is:
\begin{equation}
    \begin{split}
        \vec{a}_{\rm disk-sink}& = \frac{\vec{F}_{\rm disk}+\vec{F}_{\rm sink}}{M_{\rm disk}+M_{\rm sink}} \\
        & = \frac{M_{\rm sink}}{M_{\rm disk}+M_{\rm sink}} \vec{a}_{\mathrm{sink}} \\
        & \quad -\frac{1}{M_{\rm disk} + M_{\rm sink}} \sum_{i=1}^{N_{\rm part}} m_{i} \vec{a}_{\mathrm{disk}, i}.
    \end{split}
\end{equation}
Here, $\vec{a}_{\mathrm{sink}} = \sum G m_i \vec{r}_i/|r_i^3|$ is the point-mass-like acceleration (second equality in Eq.~\ref{eq:A1}) on the sink particle from all other particles in the simulation as evaluated by the gravity tree code in \textsc{Gadget-2}. The second term accounts for the `back reaction' on the disk itself from all other particles that feel the additional disk potential. We apply the acceleration vector $\vec{a}_{\rm disk-sink}$ in every simulation time step to advance the sink particle that marks the disk potential center.

The acceleration on the $i$-th particle ($\vec{a}_{i, \mathrm{disk}} = -\vec{a}_{\mathrm{disk}, i}$) induced by the disk potential is obtained by taking the gradient of $\Phi(R,Z)$:
\begin{equation}
    \vec{a}_{\rm disk} = -\nabla \Phi(R, Z) = -\frac{\partial \Phi(R,Z)} {\partial R} \hat{R} -\frac{\partial \Phi(R,Z)}{\partial Z} \hat{Z}.
\end{equation}
The magnitude of $\vec{a}_{i,\mathrm{disk}}$ along the radial direction is:
\begin{equation}
\begin{split}
    -\frac{\partial \Phi(R,Z)}{\partial R}
    & = -\frac{G M_{d,\ast} R}{\left[\left(\sqrt{h_{d,\ast}^2+Z^2}+R_{d,\ast}\right)^2+R^2\right]^{3 / 2}}\\
    & \quad -\frac{G M_{d,g} R}{\left[\left(\sqrt{h_{d,g}^2+Z^2}+R_{d,g}\right)^2+R^2\right]^{3 / 2}}\\
    & \quad -\frac{G M_b R}{\sqrt{R^2+Z^2}\left(\sqrt{R^2+Z^2}+r_b\right)^2},
\end{split}
\end{equation}
while the disk acceleration magnitude along the azimuthal direction is:
\begin{equation}
\begin{split}
    -\frac{\partial \Phi(R,Z)}{\partial Z}
    & = - \frac{G M_{d,\ast} \left(\sqrt{h_{d,\ast}^2+Z^2}+R_{d,\ast}\right)Z}{\sqrt{h_{d,\ast}^2+Z^2}\left[\left(\sqrt{h_{d,\ast}^2+Z^2}+R_{d,\ast}\right)^2+R^2\right]^{3 / 2}} \\
    & \quad -\frac{G M_{d,g} \left(\sqrt{h_{d,g}^2+Z^2}+R_{d,g}\right)Z}{\sqrt{h_{d,g}^2+Z^2}\left[\left(\sqrt{h_{d,g}^2+Z^2}+R_{d,g}\right)^2+R^2\right]^{3 / 2}} \\
    & \quad -\frac{G M_b Z}{\sqrt{R^2+Z^2}\left(\sqrt{R^2+Z^2}+r_b\right)^2}.
\end{split}
\end{equation}

The disk density distribution, as shown in Fig.~\ref{fig:density}, can be obtained via the Poisson equation:
\begin{equation}
    \nabla^2\Phi = \frac{1}{R}\frac{\partial}{\partial R} \left(R\frac{\partial\Phi}{\partial R} \right) + \frac{\partial^2\Phi}{\partial Z^2} = 4\pi G \rho_{\rm Disk}(R, Z).
\end{equation}

The total disk density can be expanded into three components consisting the stellar disk, gaseous disk, and the spherical bulge:
\begin{equation}
    \rho_{\rm Disk}(R, Z) = \rho_{d, \ast}(R, Z) + \rho_{d, g}(R, Z) + \rho_{\rm b}(R, Z).
\end{equation}
The stellar and gaseous disks have the same density distribution shape $\rho_d$ with different scale radii and height. The density $\rho_{d}$ can be decomposed into a radial and an azimuthal term $\rho_d = \rho_{d, R} + \rho_{d, Z}$. The radial term expands as:
\begin{equation}
    \rho_{d, R} (R, Z) = \frac{M_d\left[2\left(R_d + \sqrt{Z^2+h_d^2} \right)^2-R^2\right]}{4\pi\left[ R^2 + \left(R_d + \sqrt{Z^2+h_d^2} \right)^2 \right]^{\frac{5}{2}}}.
\end{equation}
The azimuthal term expands as:
\begin{equation}
\begin{split}
    \rho_{d, Z} (R, Z) &= \quad \frac{M_d Z^2}{4\pi\left(Z^2+h_d^2\right) \left[ R^2 + \left(R_d + \sqrt{Z^2+h_d^2} \right)^2 \right]^{\frac{3}{2}}}\\
    &\quad + \frac{M_d \left(R_d + \sqrt{Z^2+h_d^2} \right)}{4\pi\left(Z^2+h_d^2\right)^{\frac{1}{2}}\left[ R^2 + \left(R_d + \sqrt{Z^2+h_d^2} \right)^2 \right]^{\frac{3}{2}}}\\
    & \quad -\frac{M_d Z^2 \left( R_d + \sqrt{Z^2+h_d^2} \right)}{4\pi\left(Z^2+h_d^2\right)^{\frac{3}{2}}\left[ R^2 + \left(R_d + \sqrt{Z^2+h_d^2} \right)^2 \right]^{\frac{3}{2}}} \\
    &\quad -\frac{3 M_d Z^2 \left( R_d + \sqrt{Z^2+h_d^2} \right)^2}{4\pi\left(Z^2+h_d^2\right) \left[ R^2 + \left(R_d + \sqrt{Z^2+h_d^2} \right)^2 \right]^{\frac{5}{2}}} .
\end{split}
\end{equation}
The $M_d$, $r_d$ and $h_d$ constants take values either with subscript `$\ast$' for the stellar disk or subscript `g' for the gaseous disk in Section~\ref{sec:method:disk}. Finally, the bulge density profile is:
\begin{equation}
    \rho_b(R, Z) = \frac{M_b r_b}{2\pi \sqrt{R^2+Z^2} \left(r_b+ \sqrt{R^2+Z^2}\right)^3}
\end{equation}

\section{Axis ratio of the MW host halo}
\label{sec:app:axis}

\begin{figure}[!htbp]
    \centering
	\includegraphics[width=\columnwidth]{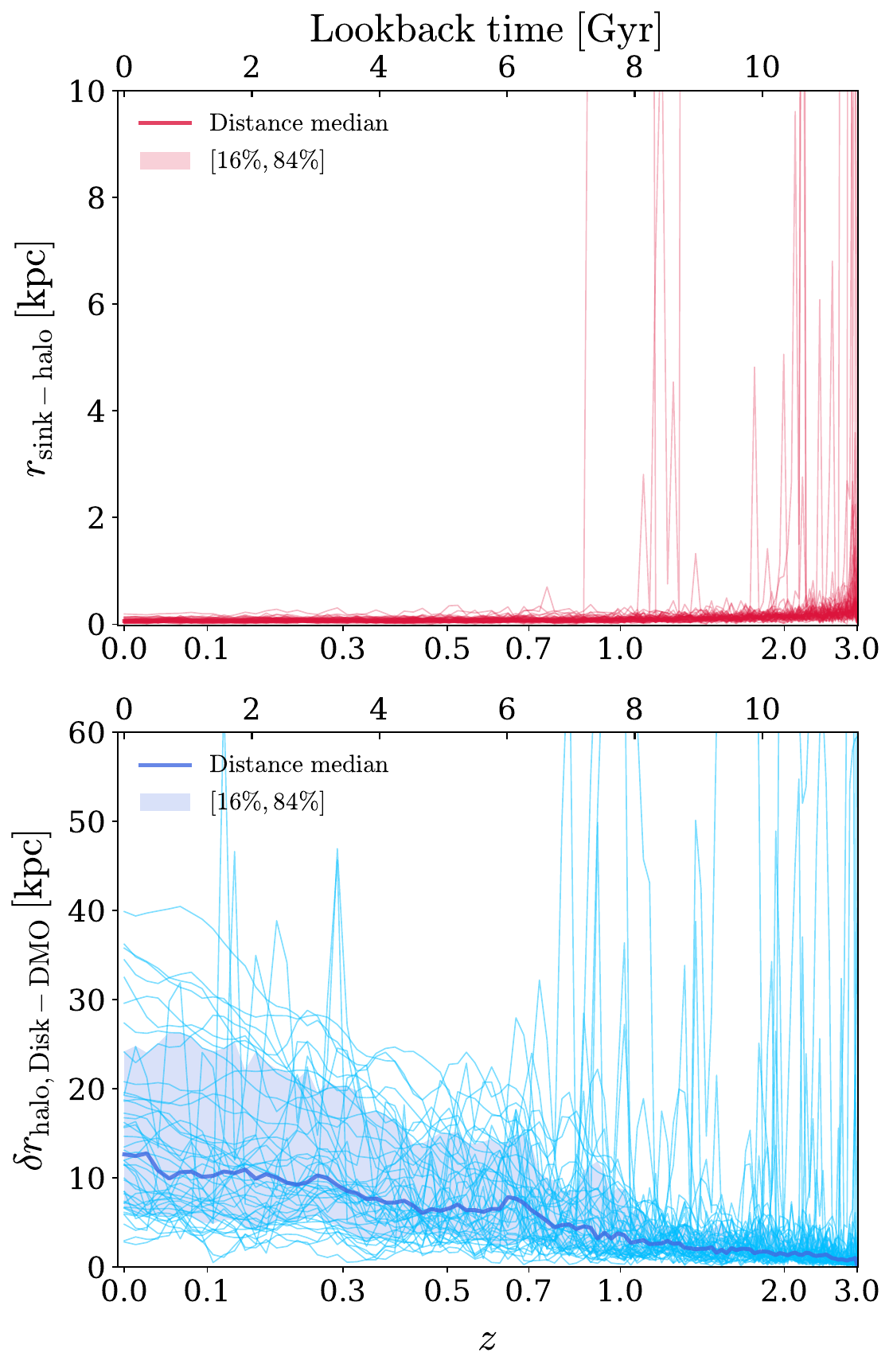}
    \caption{{Top panel}: Distance between the sink particle location and the \textsc{Rockstar}-defined halo center (potential minimum). Dynamical friction keeps  the sink at $r<1$ kpc to the halo center, with occasional out shots due to mergers. {Bottom panel}: Distance between the SymphonyMilkyWay halo center and the EDEN halo center. In each panel, the thin curves stand for individual MW-mass host. The thick solid curve is the median while the shaded region shows the $68\%$ host-to-host scatter. }
    \label{fig:traj}
\end{figure}

Dark matter density maps of the embedded disk host halos in \citet{2017MNRAS.471.1709G}
and \citet{2019MNRAS.487.4409K} show rounder shapes than their DMO counterparts due to the adiabatic contraction (Section~\ref{sec:results:mdisk}) caused by the disk potential. We observed a similar phenomenon for EDEN (Fig.~\ref{fig:map}), where hosts with larger disk-to-halo mass ratios seem to have rounder halos than hosts with lighter disks. We show the semi-medium ($b/a$) and semi-minor ($c/a$) axis ratios of each EDEN fiducial and EDEN $M_{\ast, \rm Disk}\times 2.5$ halo in Fig.~\ref{fig:axis}. The halo axis ratios were determined using the moment of inertia of all dark matter particles within $R_{\rm vir, host}$. Both the semi-medium and semi-minor axis ratios tend become larger comparing EDEN to Symphony DMO counterparts, with semi-minor axis ratios strictly increasing (all values are above 1). The increase in both axis ratios become larger with increasing disk-to-halo mass ratio, although significant scatter also exists for both correlations. 

\begin{figure*}[t!]
    \centering
	\includegraphics[width=2\columnwidth]{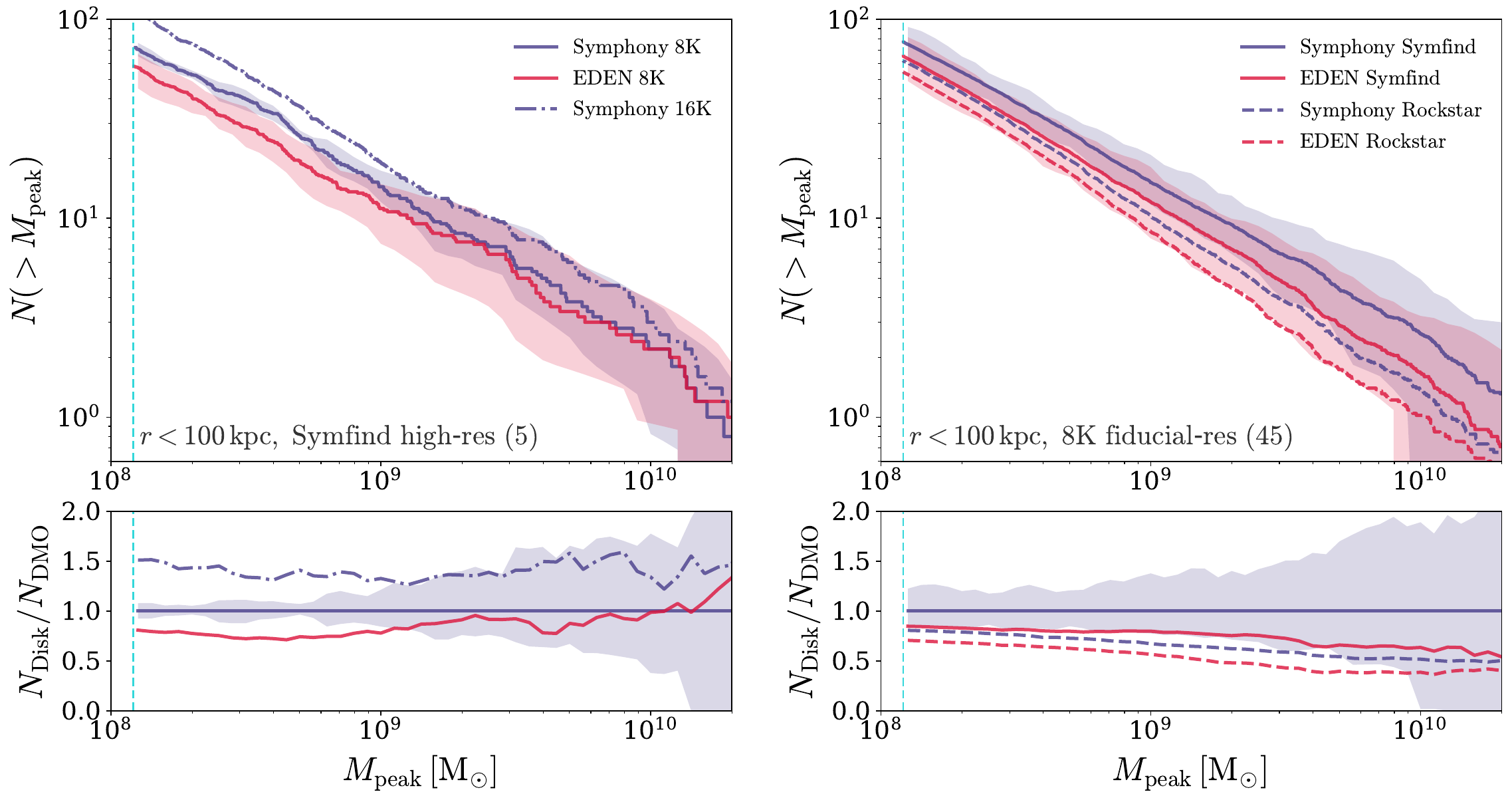}
    \caption{{Left panels:} Numerical resolution convergence tests using \textsc{Symfind} halo catalogs. This panel shows SPMF for 5 SymphonyMilkyWay (blue dotted-dashed) host halos re-simulated with $8\times$ higher numerical resolution (high-resolution, 16K) than the 45 halos presented in the main text ran at `fiducial' resolution (8K). The fiducial resolution (8K) curves in this panel also only include these 5 halos. The high-resolution DMO run contains $\sim10\%$ more subhalos at all $M_{\rm peak}$ than in fiducial-resolution, their difference being comparable to the disk effect intensity at 8K. {Right panels}: Subhalo finder comparison tests. All curves contain the entire 45 halo suites of either Symphony (blue) or EDEN (red). Solid stands for \textsc{Symfind}, which is identical to the bottom left panel of Fig.~\ref{fig:SHMF} and dashed stands for \textsc{Rockstar}. \textsc{Symfind} tracks more subhalos than \textsc{Rockstar} at all $M_{\rm peak}$, especially at the high-mass end.} 
    \label{fig:SHMF_conv}
\end{figure*}

\section{Host trajectories and subhalo infall times}
\label{sec:app:traj}

As discussed in Section~\ref{sec:discussion:disk}, the simplistic treatment of fixing the disk orientation may impact the halo spin and its 3D position by exerting a torque on the dark matter particles. In this Appendix, we show how well the sink particle is anchored to the halo center (potential minimum) and how much the added disk potential changes the spatial trajectory of the host halo. 

In Fig.~\ref{fig:traj}, we present these two distances for all 45 SymphonyMilkyWay halos. In the top panel, the median sink particle to halo center distance in the embedded disk simulations is mostly within 0.1 kpc with occasional jumps during mergers which are mostly due to temporary mis-classification of halo centers. This justifies that the sink particle approach for determining the disk potential center works reasonably well for our MW-mass halos. 

However, in the bottom panel, we observe that the distance between the halo center of the MW host halo in the DMO run starts to deviate from the Disk run since $z=3$ and reaches a median distance of 12 kpc at $z=0$. Five halos end up more than 30 kpc away from their location in the DMO run once the embedded disk potential is added, which is $\gtrsim 10\%$ of their halo $R_{\rm vir}$. The Phat-ELVIS team also kindly provided us trajectory data for one of their ten halos (private comm. James Bullock, Hyun-su Kim). It also shows a 27 kpc offset between their DMO and embedded Disk halos at $z=0$. Therefore, this systematic offset is a generic caveat for embedded disk simulations with fixed disk orientation. 

\begin{figure*}
    \centering
	\includegraphics[width=2\columnwidth]{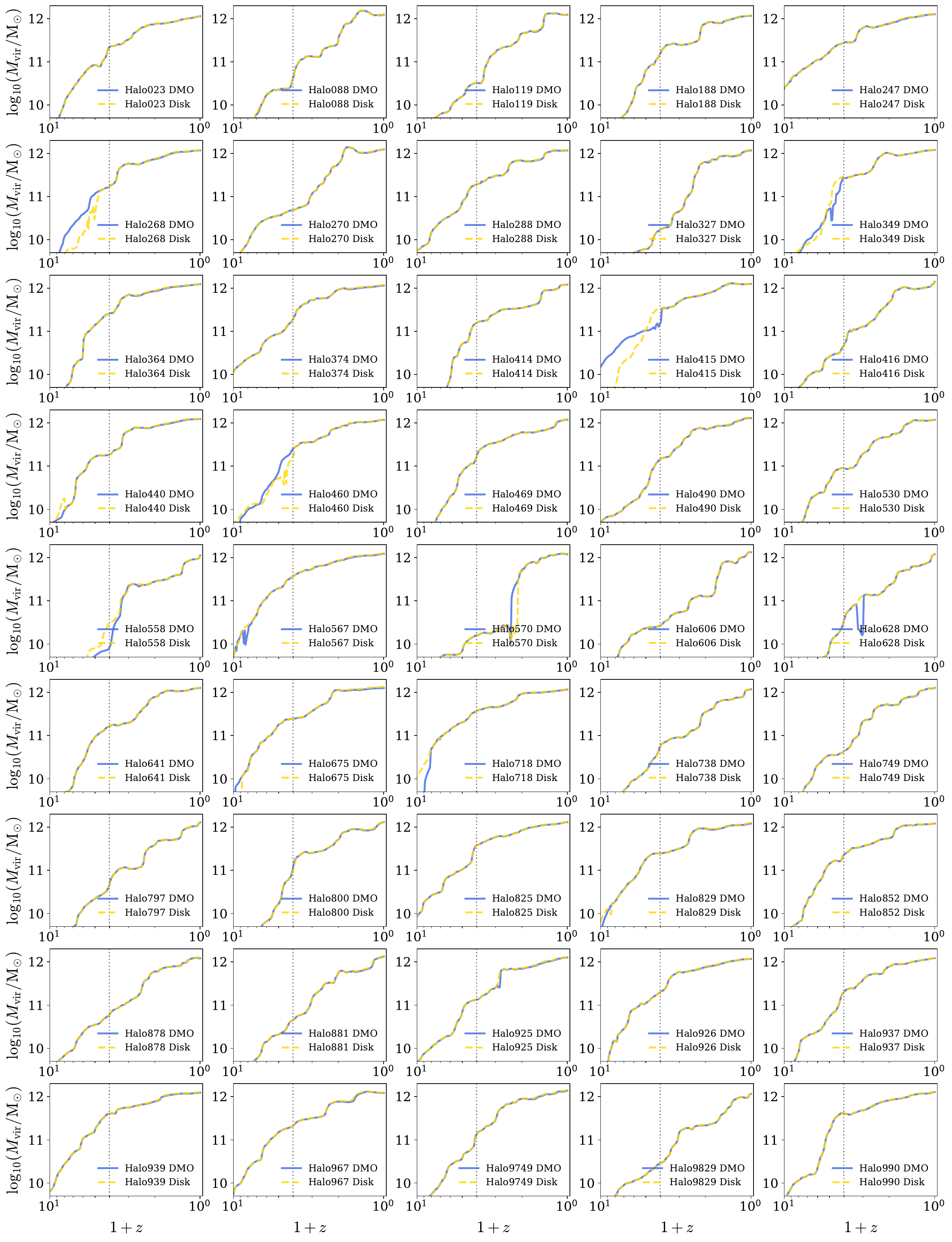}
    \caption{Halo mass accretion histories for all 45 SymphonyMilkyWay halos. The solid blue curves are the DMO runs, while the dashed yellow curves are the EDEN disk runs. The vertical dotted line in each panel marks $z=3$ where the disk potentials were seeded. Subhalo infall times and masses are in agreement between the DMO and Disk realizations for all 45 halos by comparing their mass accretion histories.}
    \label{fig:MvirH}
\end{figure*}

Nonetheless, this systematic shift in the halo location does not seem to drastically impact the subhalo masses and infall times, thus it does not significantly impact our conclusions in this paper. We show in Fig.~\ref{fig:MvirH} the mass accretion histories of every MW-mass host halo used in this work. After the addition of the disk potential at $z=3$, the mass accretion histories of the embedded-disk halos are very well aligned with the DMO halos till $z=0$, indicating that subhalo infall times and subhalo infall masses are well-conserved with the addition of the disk potential. This is expected as the shifts of the MW host halos are quite small ($\lesssim 0.1 R_{\rm vir, host}$) compared to their isolated large-scale environment as they are selected to be the largest halo within $4R_{\rm vir, host}$~\citep{2015ApJ...810...21M,2023ApJ...945..159N} of their vicinity. This also suggests that the cumulative shift in the halo center in Fig.~\ref{fig:traj} happens in a concerted manner for objects in the halo vicinity that does not impact subhalo infall times and infall masses. Our conclusions in this work therefore are not affected by the caveat of halo center offsets between the DMO and Disk runs.

\section{Resolution tests and comparison between halo finders}
\label{sec:app:res}

As mentioned in Section~\ref{sec:discussion:resolution}, the quantitative results of subhalo abundances (i.e., the subhalo count ratio) might be sensitive to the numerical resolution of the simulation. We show in Fig.~\ref{fig:SHMF_conv} how numerical resolution and the specific choice of the halo finder affect the subhalo peak mass function quantitatively.

In the left panel, we compare SPMF in \textsc{Symfind} halo catalogs for Symphony and EDEN fiducial halos at their fiducial resolution (8K) and $8\times$ higher resolution (16K). Comparing the Symphony 8K and 16K curves, more subhalos will be tracked and survive till $z=0$ at all subhalo mass scales with $M_{\rm peak}>300 m_{\rm DM}=1.2\times 10^8\mathrm{M_{\astrosun}}$. The change in SPMF due to the change in resolution is comparable to the disk effects on subhalo abundance at the fiducial 8K resolution. This indicates that the EDEN SPMF and subhalo count ratio are not fully converged under the current resolution, and that the quantitative arguments in this work on subhalo abundance suppression are sensitive to numerical resolution.  We emphasize that the qualitative conclusion of subhalo abundance suppression happening at all subhalo mass scales is unchanged with increasing resolution. We plan to re-simulate more EDEN and Symphony hosts in the future at 16K resolution.

In the right panel, we compare the SPMF at the fiducial 8K resolution for EDEN and Symphony halo catalogs constructed using \textsc{Symfind} and \textsc{Rockstar}. Within each type of subhalo finder, the EDEN embedded disk simulations consistently lose subhalos at all mass scales above the resolution limit. However, comparing between the two halo finders, we see that \textsc{Rockstar} tracks significantly less subhalos than \textsc{Symfind}, with $\sim 25\%$ less subhalos at $M_{\rm peak}\sim 10^8 \mathrm{M_{\astrosun}}$ and $\sim 50\%$ less subhalos at $M_{\rm peak}\sim 10^{10} \mathrm{M_{\astrosun}}$. It is interesting to see that the Symphony \textsc{Rockstar} curve is everywhere below the EDEN Symfind curve, meaning that the change of the subhalo finder has an even larger effect than the disk itself. These findings are consistent with the \textsc{Symfind} methodology paper~\citet{2024ApJ...970..178M} and we re-emphasize that future simulation comparisons to EDEN should be based on halo catalogs constructed with the same halo finder, \textsc{Symfind}. 

\section{Comparing to the N18 random forest model}
\label{sec:app:RF}

\begin{figure}[htbp!]
    \centering
	\includegraphics[width=\columnwidth]{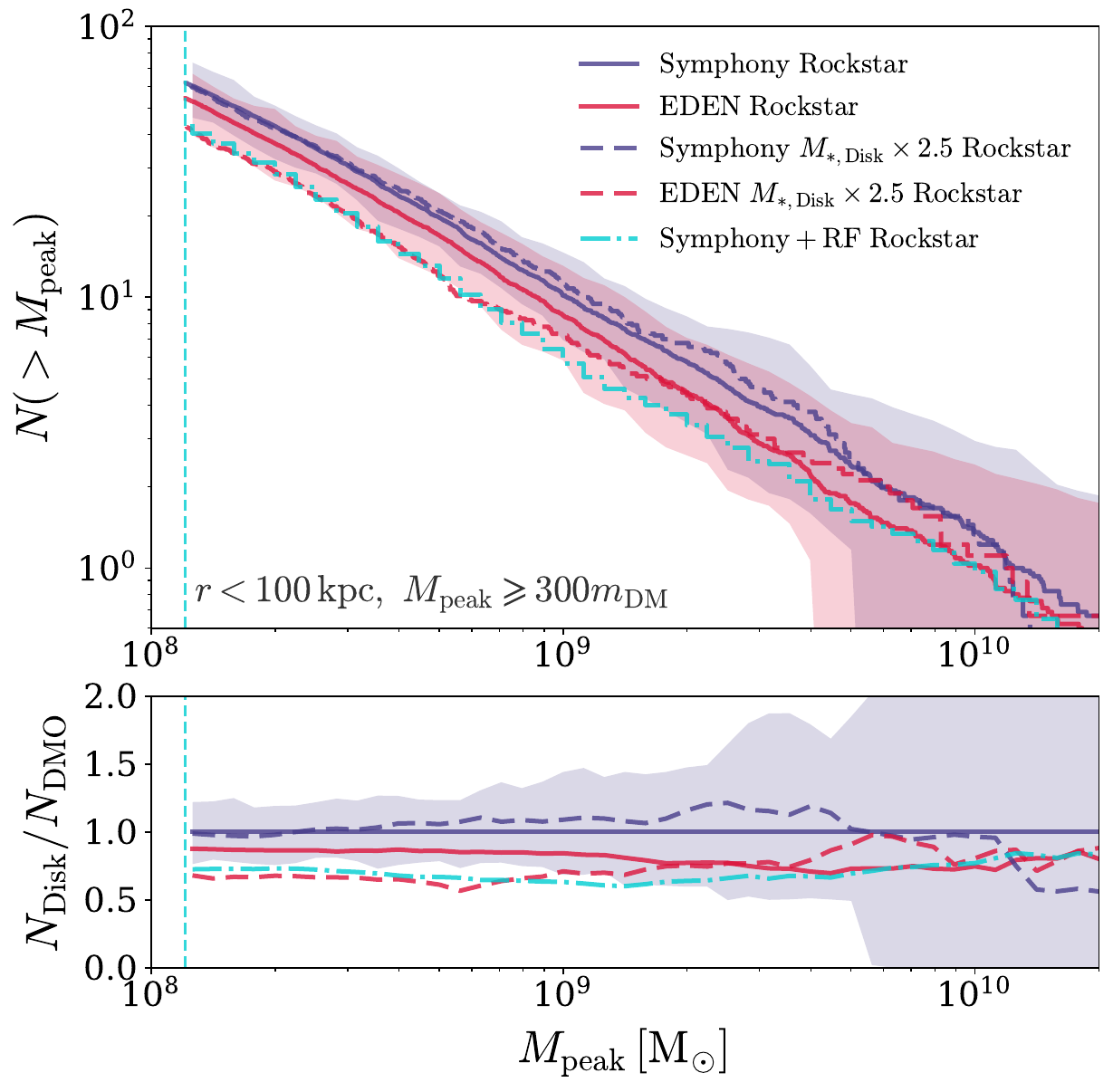}
    \caption{{Top panel:} \textsc{Rockstar} subhalo peak mass functions, compared to N18 Random Forest (RF) applied to Symphony DMO hosts. The solid curves show the Symphony DMO (blue) and EDEN fiducial (red) host, while the dashed curves show the EDEN $M_{\ast, \rm Disk} \times 2.5$ subsample and their DMO counterparts. The turquoise dotted-dashed curve is SPMF for the 45 Symphony DMO hosts after applying the RF model. {Bottom panel:} SPMF ratios relative to Symphony. The RF model subhalo abundance suppression intensity is consistent with EDEN $M_{\ast, \rm Disk} \times 2.5$ given it being trained on the \citet{2017MNRAS.471.1709G} halos with MW-like heavy disks.} 
    \label{fig:RF}
\end{figure}

To connect our model predictions to latest observational constraints on subhalo abundances, we apply the random-forest (RF) model from \citet{2018ApJ...859..129N} (N18 for short) to the  SymphonyMilkyWay DMO halos. The RF model was trained on the hydro--DMO pairs of the two halos introduced in \citet{2017MNRAS.471.1709G}, which were run using the FIRE-2 hydrodynamic simulation code~\citep{2018MNRAS.480..800H}. The model uses 5 subhalo properties (pericenter distance, pericenter scale factor, infall mass, infall scale factor, and infall maximum circular velocity) from the DMO run to predict the probability that their counterparts in the corresponding hydrodynamic simulations are lost due to tidal stripping. Since \citet{2017MNRAS.471.1709G} used \textsc{Rockstar} halo catalogs, we also apply the N18 RF model to the SymphonyMilkyWay \textsc{Rocsktar} catalogs. Given the large fraction of subhalos numerically lost track by \textsc{Rockstar} (Fig.~\ref{fig:SHMF_conv}), we caution the reader to treat this subhalo abundance suppression comparison of the RF model and actual disk potentials as a qualitatively illustration.

In \citet{2020ApJ...893...48N}, the suppression strength of this random-forest model was parameterized and constrained by observed MW satellites with abundance matching, which yielded a disk effect intensity consistent with (but with broad posteriors about) the fiducial FIRE-2 results. Following the procedures in \citet{2018ApJ...859..129N}, we multiply all mass-related properties by $(1-f_b)$ and all velocity-related properties by $\sqrt{1-f_b}$ when calculating a DMO subhalo's survival probability (we assume the cosmic baryon fraction $f_b=0.158$, \citealt{2019MNRAS.488.3143B}). 

The results of this model applied to Symphony DMO hosts are shown in Fig.~\ref{fig:RF}. The RF model applied on the full suite of 45 MW-mass halos is consistent with the SPMFs of the nine halos in EDEN $M_{\ast, \rm Disk} \times 2.5$ within $100$ kpc, except for subhalos at $M_{\rm peak} \sim 6\times 10^{9}\mathrm{M_{\astrosun}}$. The difference at this subhalo mass scale is mainly due to switching to \textsc{Rockstar} catalogs for EDEN $M_{\ast, \rm Disk} \times 2.5$, which may have introduced `merger tree' stitching errors that assigned lower mass progenitors to higher mass descendants after pericenter passage~\citep{2024ApJ...970..178M}. Nonetheless, the RF agrees well with EDEN $M_{\ast, \rm Disk} \times 2.5$  apart from this mass scale and loses significantly more lower-mass subhalos ($M_{\rm peak}\lesssim 10^9\mathrm{M_{\astrosun}}$) than EDEN fiducial. This agreement with the high-disk-mass subsample in EDEN is expected as the RF model was trained on the two \citet{2017MNRAS.471.1709G} halos with large MW-like disk masses that lead to strong subhalo abundance suppression (Fig.~\ref{fig:Nratio} top panel). 

\section{The impact of the gaseous disk}
\label{sec:app:gas}

In the EDEN disk potential setup (Section~\ref{sec:method:disk}), we have kept the `gaseous' disk potential a fixed fraction of the total disk mass throughout $z=3$ to $z=0$ (Eq.~\ref{eq:potential}). We assumed that the EDEN total disk mass follows the UM-predicted \emph{stellar} mass and did not account for any additional gas mass in order to make fairer comparisons with previous embedded disk simulations~\citep{2017MNRAS.471.1709G,2019MNRAS.487.4409K}. \citet{2017MNRAS.471.1709G} has shown that adding an additional gas disk that follows the gaseous disk growth in their hydrodynamic counterpart does not affect subhalo abundance (see their Fig. 7). They suggested that this could be due to the redshift mismatch between when the gas disk constituted a significant fraction of the baryonic mass at $z\gtrsim 1$~\citep[e.g., ][]{2021A&A...648A..25Z,2022ApJ...941L...6C} and the majority of subhalos that survive to $z=0$ falling in at $z\lesssim 1$~\citep{2015ApJ...807...49W}, diminishing the impact of the gaseous disk component.

To further verify this finding, we randomly select three hosts out of the nine high-mass subsample EDEN hosts in Section~\ref{sec:method:highmass} and re-simulate their disk potentials with an additional gaseous disk setup. We first renormalize the masses of the stellar disk $\Phi_{\ast}(R, Z)$ and the stellar bulge $\Phi_{b}(r)$ in Eq.~\ref{eq:potential} such that they sum up to the total stellar mass $M_{\ast, \rm UM}(z)$ given by UM at every snapshot (Fig.~\ref{fig:sfh}). Next, we normalize the gaseous disk potential mass in a time-varying fashion with respect to the UM-predicted stellar masses based on cosmic baryon densities predicted by the empirical gas--halo connection model \textsc{NeutralUniverseMachine}~\citep{2023ApJ...955...57G}. \textsc{NeutralUniverseMachine} is an empirical gas--halo connection model anchored on the star formation histories of the UM DR1 model~\citep{2019MNRAS.488.3143B} and matches the HI/$\mathrm{H_2}$-stellar mass functions, HI-halo mass function, molecular-to-atomic ratio, and cosmic HI/$\mathrm{H_2}$ densities from $z=6$ to $z=0$. For simplicity, we assume that the adjusted gas disk mass $M_{\rm gas}(z)$ follows a time-varying ratio with respect to the UM-predicted stellar mass $M_{\ast, \rm UM}(z)$:
\begin{equation}
    \label{eq:gas_ratio}
    \frac{M_{\rm gas}(z)}{M_{\ast, \rm UM}(z)} = \frac{1.36\rho_{\rm HI}(z)+\rho_{\rm H_2}(z)}{\rho_{\ast}(z)},
\end{equation}
where $\rho_{\rm HI}(z)$, $\rho_{\rm H_2}(z)$, and $\rho_{\ast}(z)$ are the cosmic atomic hydrogen, molecular hydrogen, and stellar mass densities predicted by \textsc{NeutralUnvierseMachine} (their Fig. 11), and the 1.36 factor accounts for elements heavier than Hydrogen.

\begin{figure}[htbp!]
    \centering
	\includegraphics[width=\columnwidth]{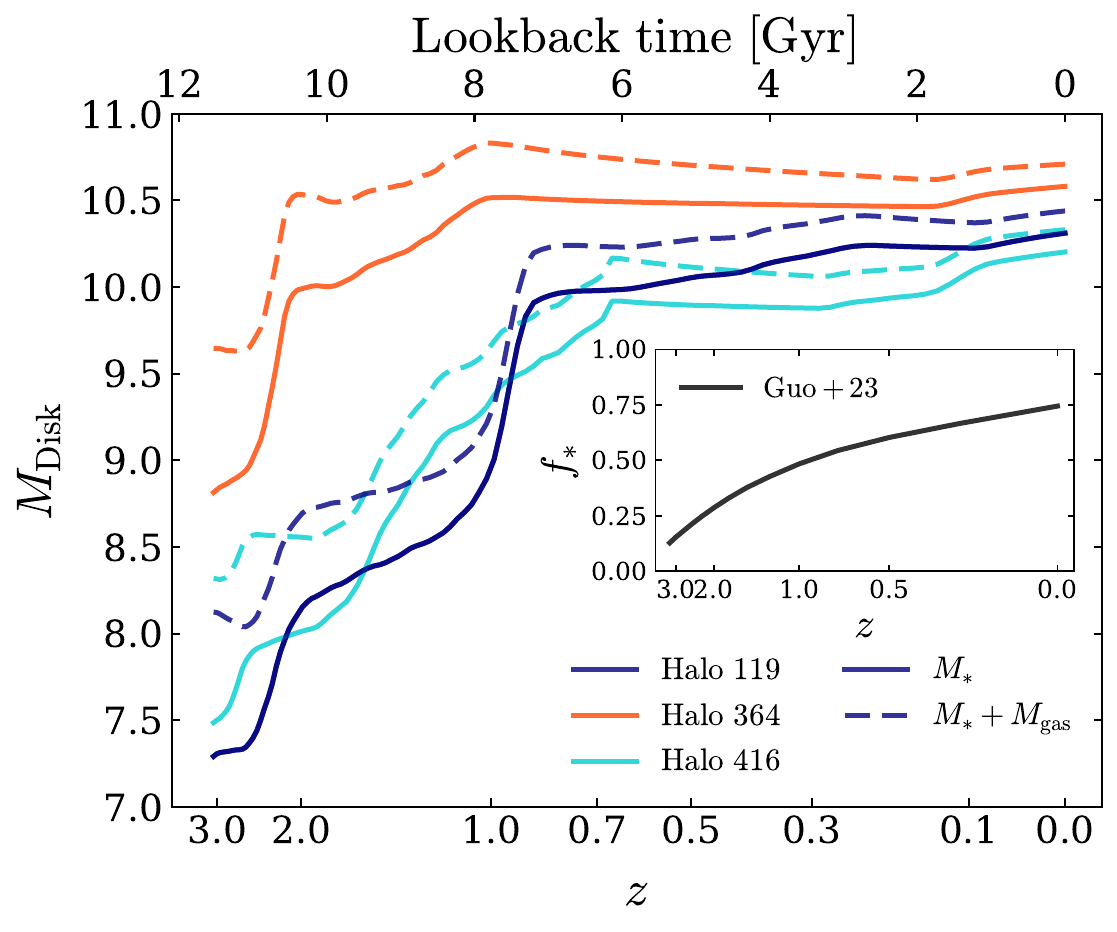}
    \caption{{\it Main panel:} Total disk mass history of the fiducial UM-predicted stellar mass (solid) and the gas-adjusted disk masses that follow time-varying gas mass ratios according to Eq.~\ref{eq:gas_ratio} (dashed). The slight decreasing trend in parts of the stellar mass histories is a manifestation of UM accounting for massive stars' deaths over time. {\it Inset:} The implied universal stellar mass fraction $f_{\star}$ is inferred from the \textsc{NeutralUniverseMachine}-predicted cosmic HI, $\rm{H_{2}}$, and stellar mass densities (see Fig. 11 in \citealt{2023ApJ...955...57G}.)}
    \label{fig:gasH}
\end{figure}

This is a limiting model, and for simplicity does not account for the variations in gas fraction across different halo mass scales. Because of this, we are likely to be overestimating the impact of gas disks since a MW-mass halo will often have less gas content than the \citet{2023ApJ...955...57G} average. However, given the modeling uncertainty in the evolution of gas content, testing a slightly extreme model can provide an upper bound estimate on the subhalo abundance suppression levels without loss of generality. We show the fiducial UM-predicted total disk mass and the gas-adjusted total disk mass histories along with the stellar mass fraction of the adjusted disks according to Eq.~\ref{eq:gas_ratio} for the three selected EDEN hosts in Fig.~\ref{fig:gasH}.

\begin{figure}[htbp!]
    \centering
	\includegraphics[width=\columnwidth]{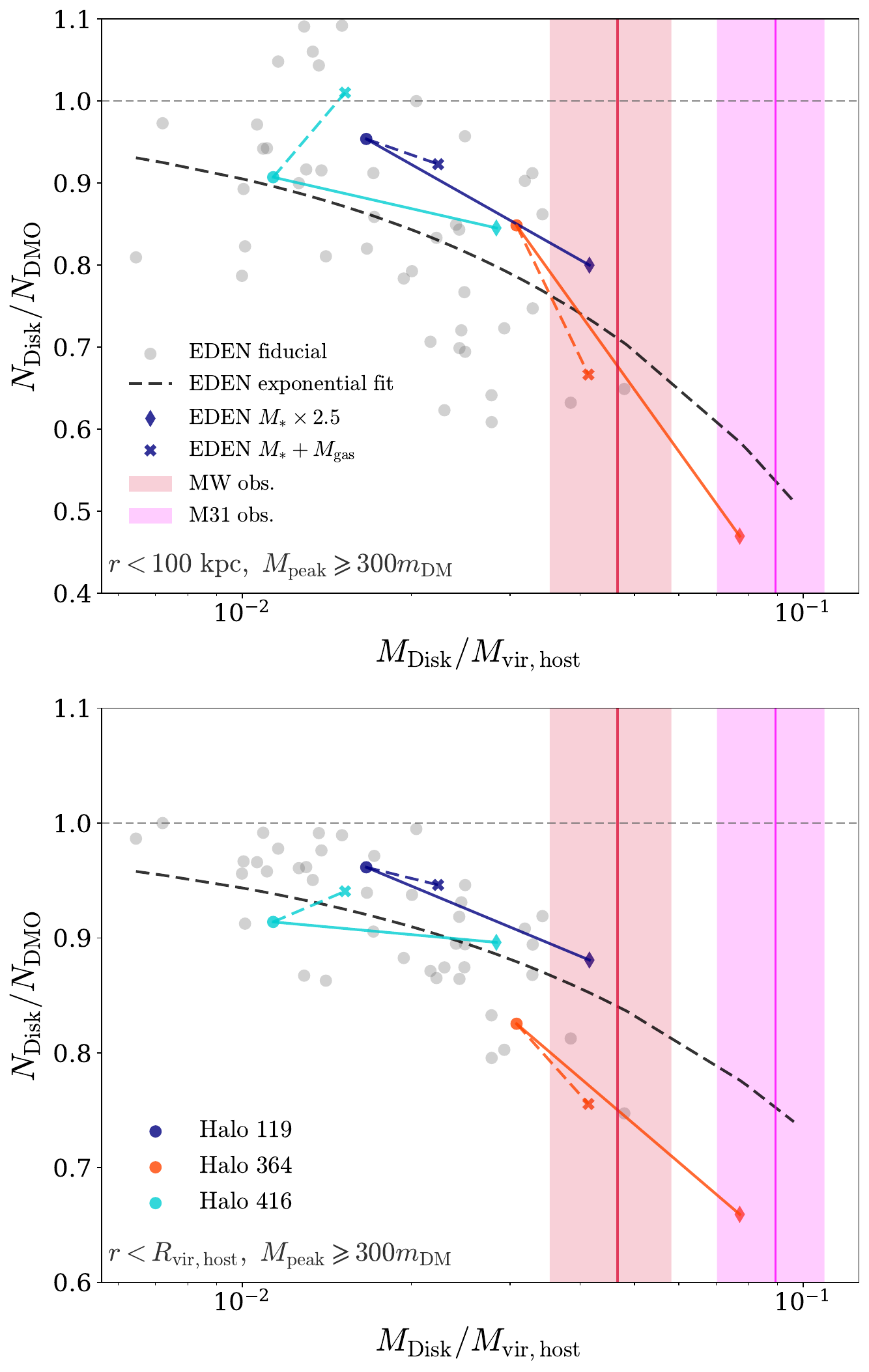}
    \caption{Subhalo count ratios within 100 kpc (top) and $R_{\rm vir, host}$ (bottom) for the three adjusted gas-disk EDEN re-simulations (crosses). The gas-adjusted disk re-simulations are connected to their fiducial EDEN counterparts using dashes lines. The $M_{\ast}\times2.5$ high stellar-mass counterparts are also shown for reference (diamonds, connected to fiducial EDEN using solid lines). The three gas-adjusted re-simulated halos shown in the same color coding as their mass history curves in Fig.~\ref{fig:gasH}. The adjusted gas disk with higher gas masses at earlier redshift has a smaller impact on $N_{\rm Disk}/N_{\rm DMO}$ compared to the mass normalization of the stellar disk.}
    \label{fig:Nratio_gas}
\end{figure}
 
We show in Fig.~\ref{fig:Nratio_gas} the subhalo count ratios of the three re-simulations of EDEN hosts with gas-adjusted disk potentials. We compare the subhalo abundance suppression ratio $N_{\rm Disk}/N_{\rm DMO}$ of the gas-adjusted re-simulations with their fiducial EDEN runs and their $\times 2.5$ stellar disk mass re-simulations. Two of the three hosts we re-simulate with the adjusted gas disk potential show reduced $N_{\rm Disk}/N_{\rm DMO}$ compared to their fiducial EDEN counterparts, while one of them shows increased $N_{\rm Disk}/N_{\rm DMO}$. We summarize that:
\begin{itemize}
\item The overall effect of adding a gas disk is similar to moving along the fitted $N_{\rm Disk}/N_{\rm DMO}$ relation, with the $z=0$ baryonic mass being the primary driver of subhalo abundance suppression without needing to differentiate between how much mass is in stars versus gas.

\item There is scatter in the amount of subhalo suppression/boost along the fitted relation. We cannot rule out that this scatter has some underlying physical cause (i.e., additional contraction by the gas disk) which needs a larger sample to verify. However, we note that structure formation is chaotic in nature and the gas disk could trigger a chaotic response that significantly changes subhalo infall times. Because of this we urge caution in over-interpreting individual hosts.

\item Given that the stellar and gaseous disks evolve differently, one explanation to the gas disk insensitivity could be that most subhalos fall in at  $z\lesssim 1$  when the stellar disk starts to dominate ($f_{\ast} >50\%$, Fig.~\ref{fig:gasH}).
\end{itemize}

In all, these findings are consistent with the findings in \citet{2017MNRAS.471.1709G}, where the gas disk is a less decisive factor of subhalo abundance suppression due to disk tidal stripping. This also makes our choice of a fixed-gas-ratio choice for the disk model more justifiable in Section~\ref{sec:method:disk}. Although the gas disk appears to be less decisive, our simplistic gas disk model and limited sample size do not rule out a more detailed study in the impact of the gaseous disk from being fruitful. In the future, we could explore the impact of the gas disk with higher-resolution simulations on a larger sample of hosts to further nail down systematics. 

\end{document}